\documentclass[preprint2]{aastex62}

\usepackage{rotating}

\usepackage{graphicx}
\usepackage{epstopdf}
\usepackage{natbib}
\usepackage{xspace}
\usepackage{color}
\usepackage{amsmath}

\newcommand{\kms}     {\,km\,s$^{-1}$\xspace}

\newcommand{\mjy}     {\,mJy\,beam$^{-1}$\xspace}

\newcommand{\msun}    {\,M$_{\sun}$\xspace}

\newcommand{\lsun}    {\,L$_{\sun}$\xspace}

\newcommand{\co}      {CO\,(2-1)\xspace}

\newcommand{\dechms}[4]{$#1^{\rm h}#2^{\rm m}#3\mbox{$^{\rm s}\mskip-7.6mu.\,$}#4$} 
 
\newcommand{\decdms}[4]{$-#1^{\circ}#2'#3\mbox{$''\mskip-7.6mu.\,$}#4$}

\shorttitle{The molecular environment of DO\,Tauri}
\shortauthors{Fern\'andez L\'opez et al.}
\begin{document}

\title{A ringed pole-on outflow from DO\,Tauri revealed by ALMA}

\author{Manuel Fern\'andez-L\'opez}
\affiliation{Instituto Argentino de Radioastronom\'\i a (CCT-La Plata, CONICET; CICPBA), C.C. No. 5, 1894, Villa Elisa, Buenos Aires, Argentina}
\email{manferna@gmail.com}

\author{Luis A. Zapata}
\affiliation{Instituto de Radioastronom\'\i a y Astrof\'\i sica, Universidad Nacional Aut\'onoma de M\'exico, P.O. Box 3-72, 58090, Morelia, Michoac\'an, M\'exico}

\author{Luis F. Rodr\'\i guez}
\affiliation{Instituto de Radioastronom\'\i a y Astrof\'\i sica, Universidad Nacional Aut\'onoma de M\'exico, P.O. Box 3-72, 58090, Morelia, Michoac\'an, M\'exico}

\author{Mar\'\i a M. Vazzano}
\affiliation{Instituto Argentino de Radioastronom\'\i a (CCT-La Plata, CONICET; CICPBA), C.C. No. 5, 1894, Villa Elisa, Buenos Aires, Argentina}

\author{Andr\'es E. Guzm\'an}
\affiliation{National Astronomical Observatory of Japan, National Institutes of Natural Sciences, 2-21-1 Osawa, Mitaka, Tokyo 181-8588, Japan}

\author{Rosario L\'opez}
\affiliation{Departament d’Astronomia i Meteorologia (IEEC-UB), Institut de Ci\`{e}ncies del Cosmos, U. de Barcelona, Mart\'\i i Franqu\`{e}s 1, E-08028 Barcelona, Spain}

\begin{abstract}
We present new ALMA Band\,6 observations including the \co line and 1.3\,mm continuum emission from the surroundings of the young stellar object DO\,Tauri. 
The ALMA CO molecular data show three different series of rings at different radial velocities. These rings have radii around 220\,au and 800\,au. We make individual fits to the rings and note that their centers are aligned with DO\,Tauri and its optical high-velocity jet. In addition, we notice that the velocity of these structures increases with the separation from the young star. We discuss the data under the hypothesis that the rings represent velocity cuts through three outflowing shells that are possibly driven by a wide-angle wind, dragging the environment material along a direction close to the line of sight (i$=19\degr$). We estimate the dynamical ages, the mass, the momentum and the energy of each individual outflow shell and those of the whole outflow. The results are in agreement with those found in outflows from Class\, II sources. We make a rough estimate for the size of the jet/wind launching region, which needs to be of $\lesssim 15$\,au. We report the physical characteristics of DO\,Tauri's disk continuum emission (almost face-on and with a projected major axis in the north-south direction) and its velocity gradient orientation (north-south), indicative of disk rotation for a 1-2\msun central star. Finally we show an HST [SII] image of the optical jet and report a measurement of its orientation in the plane of the sky.
\end{abstract}


\section{Introduction} \label{sec:intro}
In the heart of the Taurus molecular cloud lies the young stellar object DO\,Tauri, 
located at 139.4$^{+1.0}_{-0.9}$\,pc from Earth \citep[as measured by the Gaia\,DR2 catalogue][]{2018Gaia}. It has an spectral type M0.3 with an effective temperature of 3830\,K and a stellar luminosity of 0.2-1.2\lsun \citep{1995Kenyon,2014Herczeg}. From these properties and the assumed theoretical evolutionary tracks, the stellar mass is estimated in 0.5-0.7\msun and its age, in 0.8-9$\times10^6$\,years \citep{2002Kitamura,2002Palla,2014Herczeg}. DO\,Tauri is one of the T\,Tauri stars with highest accretion and wind ejection rates \citep{1995Hartigan,1998Gullbring,2017AlonsoMartinez}, with estimates ranging two orders of magnitude around an average value of about $\dot{M}_{acc}=10^{-7}$\msun~yr$^{-1}$ for the accretion rate and a 1\% of that value for the wind ejection rate. These large values indicate that this source is among the most active accreting Class II stars in Taurus.
DO\,Tauri is surrounded by a dusty disk discovered by \cite{1990Beckwith} using IRAM~30m observations at 1.3\,mm. \cite{1995Koerner2} detected it at a range of centimeter and millimeter wavelengths using OVRO and VLA observations. The dusty disk has not being clearly resolved until this year, and its size, inclination and position angle were uncertain for a long time \citep[][]{1995Koerner2,2002Kitamura,2015Kwon,2017Tripathi}. Recently, \cite{2019Long} resolved the dusty disk (with a 95\% radius of $0\farcs25$), reporting a low inclination of $28\degr\pm0.3\degr$ and a position angle (PA) of $170\degr\pm1\degr$. Its mass has been estimated to be  in the range 0.004-0.014\msun \citep{2002Kitamura,2005Andrews,2015Kwon}. \cite{1995Koerner1} observed an asymmetric \co profile toward the disk position showing high-velocity blue-shifted emission that did not match the disk rotation expectations. They concluded that the CO was not only tracing the disk but possibly part of an outflow. The morphology of the CO emission did not exactly match the expectations for the emission of an outflow either, suggesting the possibility of circumstellar infalling motions.
The disk is polarized at NIR wavelengths, showing a scattering position angle of 170$\degr$-180$\degr$ \citep{1982Bastien,1985Bastien,1989Tamura}, or a more recent value of $166\degr\pm1\degr$ inside a $6\arcsec$ region \citep{2009Pereyra}. At high-angular resolution, both an H-band Subaru coronagraphic and an optical STIS Hubble images, reveal an asymmetric arc-like structure about $2\arcsec-3\arcsec$ north of the star, running from PA$=45\degr$ to PA$=320\degr$ and the presence of a faint (but doubtful) companion at $\sim3\farcs5$ \citep{2008Itoh}. The interpretation of the former arc structure is complicated, although it could be related to the high-velocity bipolar jet driven by DO\,Tauri, found through the study of the [SII] forbidden line spectra \citep{1994Hirth}.
The optical jet (associated with the object HH\,230) reaches larger velocities in the red-shifted part of the spectra ($+220$\kms against $-90$\kms) and has a total size of about 8$\arcsec$ (although its optical image has never been published). The PA of its blue-shifted lobe is 250$\degr\pm10\degr$ \citep{1997Hirth}, roughly perpendicular to the NIR polarized scattered emission from the dust.

At larger scales (thousands of au), DO\,Tauri is associated with an optical arc-like reflection nebula extending northwest.
At about $91\arcsec$ east of DO\,Tauri it is located another classical T-Tauri star: the triple system HV\,Tauri. Both young stars are connected by a bridge structure preferentially seen through Herschel observations at 100\,$\mu$m and 160\,$\mu$m \citep{2013Howard}. This bridge (or common envelope) has been recently proposed as the imprint of a tidal tail structure produced by the disintegration of a putative multiple system formed by DO\,Tauri and the current HV\,Tauri triple system, about 0.1\,Myr ago \citep{2018Winter}.

In this work we analyze new Atacama Large (sub)Millimeter Array (ALMA) archive observations toward DO\,Tauri at 1.3\,mm and focus mainly on the analysis of the large-scale molecular emission from the pole-on outflow associated with this young star. In Section \ref{sec:obs} we describe the observations and explain the calibration procedure. Section \ref{sec:results} presents the results obtained from the continuum and CO emission associated with the disk, the optical jet observed by the Hubble Space Telescope (HST) and the extended \co emission associated with the outflow. We discuss our findings in Section \ref{sec:dis} and present an outline with the main ideas extracted from this work in Section \ref{sec:conclusions}. We also include a table summarizing our fit to the outflow emission in the Appendix.

\section{Observations} \label{sec:obs}
\subsection{ALMA observations}
We present ALMA Cycle 4 archive observations (Project 2016.1.01042.S; P.I. Claire Chandler) toward the surroundings of the young stellar object DO\,Tauri. During the observations carried out on 28 November 2016, 47 antennas (with baselines from 16\,m to 626\,m) were on duty and the weather was good (average precipitable water vapor column 2.05\,mm and average system temperature $\sim120$\,K) to operate at Band\,6. The array was pointed at 
$\alpha_{J2000.0}$ = \dechms{04}{38}{28}{600}, and $\delta_{J2000.0}$ = \decdms{26}{10}{49}{20} 
and the primary beam of the 12\,m antennas corresponds to about $26\arcsec$ at the observing frequency. Time on source was 1.22\,minutes.  

The correlator setup included eight spectral windows comprising about 6800\,MHz, which we used to generate a continuum image after removing the emission from the line channels (mainly from the bright \co transition). The spectral windows were centered at 232.366\,GHz (spw0), 216.994\,GHz (spw1),231.193\,GHz (spw2), 230.703\,GHz (spw3),  
218.787\,GHz (spw4), 219.256\,GHz (spw5),\\ 
219.725\,GHz (spw6) and 220.194\,GHz (spw7). For our study we are particularly interested in the spectral window 3, which contains the \co emission at a spectral resolution of 0.488\,MHz (about 0.63\kms at 230.538\,GHz).

The ALMA data were calibrated using the data reduction scripts provided by ALMA for the Common Astronomy Software Applications (CASA) package \citep{2007McMullin}, which we use in its version\,4.7.0. The atmospheric phase noise was reduced using radiometer measurements. The quasar J0510+1800 was used as bandpass and absolute flux calibrator, assuming a model with a flux density of 2.139\,Jy at 232.366\,GHz and a millimeter spectral index of $-0.393$\footnote{Here we define the spectral index $\alpha$ as $S_{\nu}\propto \nu^{\alpha}$, where $S_{\nu}$ is the flux density and $\nu$ the observed frequency.}, produced by the ALMA monitoring of quasars. Typically, the uncertainty in the ALMA flux measurements is estimated to be of order of $\sim10$\% for Band\,6. The quasar J0438+3004 was used to track and correct the phase variations. 

After the standard calibration the continuum data was imaged using CASA version\,5.4.0. We set the robust parameter to 0.5 and obtain a synthesized beam of $0\farcs71\times0\farcs47$ (PA$=149\degr$). We made three iterations of self-calibration reducing the final rms of the image by 60\% (from 0.2\mjy to 0.07\mjy) and improving the signal-to-noise ratio from 80 to 350. The final solutions found during the continuum self-calibration stage were applied to the \co spectral window and a velocity cube was constructed using robust$=0.5$. This yielded an image with a synthesized beam of $0\farcs78\times0\farcs53$ (PA$=145\degr$) at the native velocity resolution of 0.63\kms. The rms of the velocity cube in one channel is about 13\mjy (0.72\,K).

To better spatially resolve DO\,Tauri's dusty disk we construct the final continuum image using a second day of observations within the same ALMA project 2016.1.01042.S, taken on 2 August 2017 with the array in a more extended configuration (with baselines from 32\,m to 2620\,m). These data have higher-angular resolution and therefore are not as sensitive as the former observations in tracing the molecular extended emission. In the second track ALMA had 46 available antennas and the weather conditions were better, yielding a median system temperature of about 70\,K. The correlator setup and the pointing were similar to the first track of observations. J0510+1800 was used as the bandpass calibrator and J0423-0120 as the flux calibrator (assuming it has 0.835\,Jy at 223.670\,GHz and a millimeter spectral index -0.437, values extracted from the ALMA monitoring of quasars). The phase was corrected using intertwined observations of J0426+2327. The time spent on DO\,Tauri was 3.7 minutes and after three rounds of self-calibration we end up with a signal-to-noise ratio enhancement rising from 85 to more than 400. The synthesized beam of the resulting image is $0\farcs17\times0\farcs12$ (PA$=14\degr$) when using robust 0.5, and the final rms noise level is 0.06\mjy (or 0.067\,K, see Figure \ref{fdisk}). 

In the following we use the high angular resolution data set for imaging and analyzing the continuum emission due to its finest resolution. For imaging and analyzing the molecular gas emission we prepared a dataset combining both configurations, which yields a synthesized beam of $0\farcs21\times0\farcs16$ (PA$=7\degr$) and an rms noise level of 4.6\mjy (or 3.1\,K) as measured in the empty channels of the velocity cube.

\subsection{HST archive observations}

To further investigate the optical jet of DO\,Tauri we downloaded the available images of this source in the HST Legacy Archive (P.I. C. Dougados; HST proposal ID\,8215, \textit{New clues to the ejection process in young stars: Forbidden line imaging of TTauri microjets}). The data comprises forbidden line observations, in particular, we took two images.
The first one was obtained in a deep (about 2\,hours) integration using a narrowband filter (47\,\AA{} bandwidth) centered at 6732\,\AA{} (the emission of the [SII] 6717\AA{} and 6731\AA{} lines, characteristic of optical jets, are then included in the imaged wavelength range). The other image was taken with a filter of 483\,\AA{} bandwidth centered at 5483\,\AA{}, which does not include the characteristic emission lines, thus being useful to map the continuum emission. Figure \ref{fjet} present these HST data from which we could measure the position angle of the optical jet.

\section{Results} \label{sec:results}
\subsection{The disk around the young star DO\,Tauri} \label{sec:disk}
The main focus of this paper is on the study of the large scale molecular gas emission near DO\,Tauri, but we also report the characteristics of the disk dust emission, spatially resolved here, and its associated \co emission \citep[see also the recently reported ALMA results by][analogous to these simultaneously derived here]{2019Long}. The disk is detected at 1.3\,mm continuum (Figure \ref{fdisk}) with an intensity peak of 24.55$\pm0.06$\mjy and a flux density of 116.8$\pm0.1$\,mJy, consistent with the 125.4$\pm3.5$\,mJy reported by \cite{2015Kwon} at a similar frequency. These authors derived a disk mass of 0.014$\pm0.001$\,\msun by using a sophisticated disk modeling. We have fit DO\,Tauri's disk emission in the uv-plane after using a 2D-Gaussian model implemented in the task uvmodelfit in CASA. The disk is centered at ($\alpha$, $\delta$)$_{J2000}$ = (\dechms{04}{38}{28}{5935},\decdms{26}{10}{49}{272}) with a positional uncertainty of $0\farcs001$. The estimated major and minor axes of the disk are $0\farcs3447\pm0\farcs0004\times0\farcs3251\pm0\farcs0005$ (these are statistical uncertainties). Note that this is roughly a circular disk, with a $47$\,au radius and the uncertainty of its position angle and inclination may be larger than that provided by the statistical procedure. Therefore, we estimate the uncertainty of the disk PA and inclination as the difference with the same quantities derived by \cite{2019Long}. For a circular disk, this fit implies an inclination of $19\degr\pm9\degr$ with respect to the plane of the sky. Plus, its derived PA is $150\degr\pm20\degr$, not exactly perpendicular to the optical jet axis (with PA$=260\degr\pm1\degr$, see section \ref{sec:jet}), for it is deviated by about $20\degr$ from this configuration. Let us note here that an independent 2D-Gaussian fit to the continuum emission in the image plane, gives a similar result but with a different PA of $175\degr\pm6\degr$. Taking this latter result and considering the uncertainties, the orientation of the disk is consistent with being perpendicular to the jet direction.

The \co molecular image of Figure \ref{fdisk} (left panel) shows the emission integrated in the velocity ranges [0.6,5.0]\kms (blue-shifted) and [6.9,12.7]\kms (red-shifted) emission. A north-south velocity gradient is evident at the disk position. We measured the position angle of the blue-/red-shifted velocity gradient in the disk. The PA between the blue- and red-shifted peaks of the \co emission (which are apart $0\farcs22$, i.e., the size of one beam approximately) is $1\degr\pm4\degr$. This PA does not coincide the disk PA derived from the Gaussian fit to the continuum emission implemented in the uv-plane. We note that our PA measurement using the CO emission could be affected by contaminating extended emission most present at blue-shifted velocities. In any case, this PA is not exactly perpendicular to the jet axis. If real, the position angle discrepancies can be the result of a tilt or warp in the disk, but only through new more sensitive and higher angular resolution observations can the PA be more accurately measured.
Figure \ref{fdisk} (right panel) shows a position-velocity diagram along the velocity gradient of the DO\,Tauri's CO disk. The molecular emission is separated in two parts by a cloud absorption feature centered at 5.7\kms. Close to the central star (zero offset) the emission is at larger absolute velocities, which steeply decreases with the distance to the star. This is suggestive of a Keplerian rotation pattern with a velocity gradient of about 3-4\kms between the red- and blue-shifted emission at a fiducial radius of about $0\farcs3$ (about 40~au). The presence of extended emission and the spectral resolution of the current data hampers a more accurate measurement of this gradient, which we very roughly estimate to imply a dynamical mass ranging 1.1-1.9\msun (M=V$^2_{rad}$~r/(sin$^2$(i)~G, using i$=19\degr$), see also curves in the position-velocity diagram of Fig. \ref{fdisk}). At this point, the detection of the disk in the CO images and pv-diagrams is slightly speculative, and it does not merit more sophisticated modeling. Confirming and characterizing the disk will probably require to make isotope $^{13}$CO and/or C$^{18}$O observations, that are less confused with outflow emission. 

\subsection{The optical jet seen by the HST} \label{sec:jet}
Microjets are usually associated with quite evolved young stars such as Class II objects. 
From an observational point of view, this implies: 
(a) a near-infrared jet counterpart is not usually detected (H$_2$ emission); 
(b) the jet emission is short as compared with classical optical jets, consisting of just a
few knots close to its driving source; 
(c) the driving source can be detected at optical wavelengths. 
(b) and (c) cause difficulties to detect microjets, being the dominant problem the pollution of the jet emission
by the emission from the star and from the light scattered by the surrounding material. This is specially critical in deep (long time exposure) images, inevitable to detect the jet emission.

\begin{figure*}[!htb]
\minipage{0.48\textwidth}
  \includegraphics[width=\linewidth]{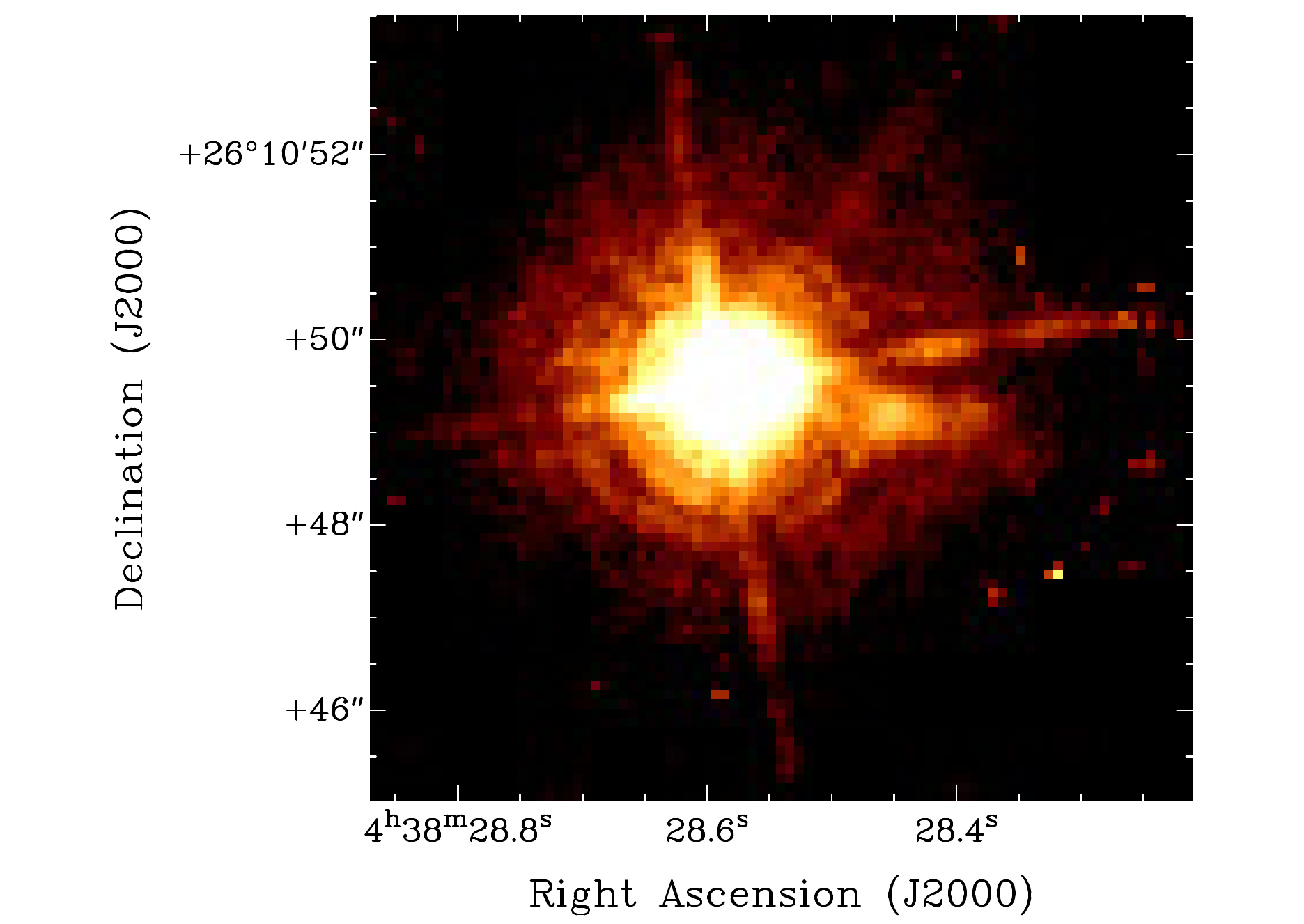}
\endminipage\hfill
\minipage{0.48\textwidth}
  \includegraphics[width=\linewidth]{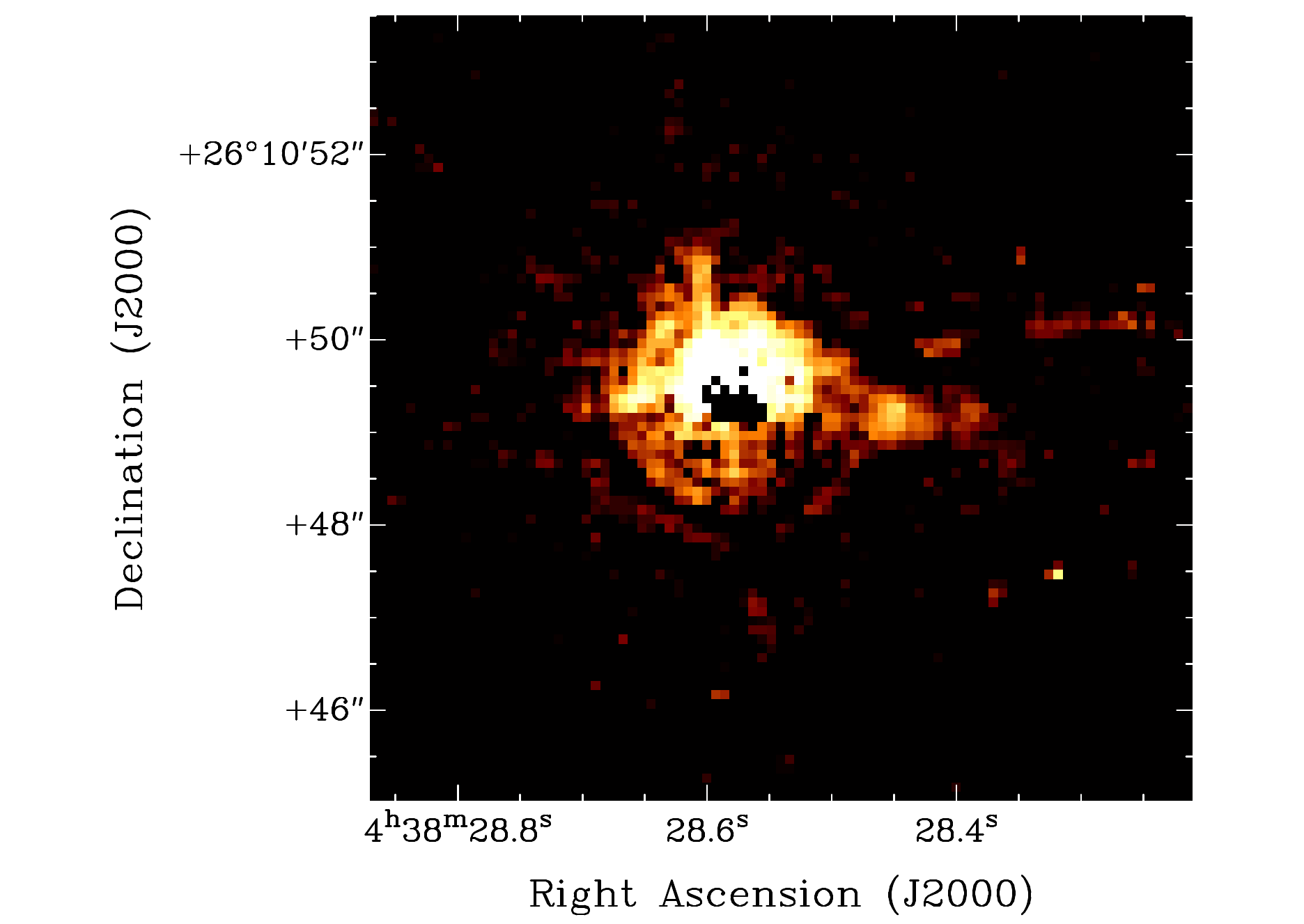}
\endminipage\hfill
  \caption{\textbf{Left:} HST image toward DO\,Tauri taken with a narrowband filter comprising the emission from the [SII] 6717\,\AA{} and 6731\,\AA{} forbidden lines. Apart from the spikes due to saturation, there is an evident collimated structure due southwest of the central source: the microjet.  
   \textbf{Right:} HST [SII] filter image (f673n filter) after continuum subtraction (f547m filter). Some bright pixels appear northeast of the central star, which could belong to the counterjet.}
  \label{fjet}
\end{figure*}

Figure \ref{fjet} shows the [SII] optical emission from the microjet in the HST images of DO\,Tauri. Although difficult to visualize due to the contamination from the source emission and the spikes produced by the saturation of the image, the emission from the microjet is clear west of the young stellar emission. The maximum of the emission of the jet has a PA of $260\degr\pm1\degr$ with respect to the peak emission toward the central object, and is $1\farcs9\pm0.1$ (270\,au) away from it. In addition, Fig. \ref{fjet} reveals two possible knots further from the protostar due southwest. These two knots are not exactly aligned with the PA of the inner jet ($260\degr$), which may suggest it is wiggling. However, a more clear detection of these knots is needed to confirm the hints of wiggling.
The measured PA of the inner part of the optical jet is roughly consistent with the previously value of $70\degr\pm10\degr$ (or its corresponding $250\degr\pm10\degr$) reported by \cite{1997Hirth}, but the difference ($10\degr$) is of the order of the uncertainty in Hirth's et al. long slit spectra measurements.  

In an attempt to look for the counterpart of the microjet we subtract the continuum emission contribution to the line emission image (right panel in Fig. \ref{fjet}). The resulting image still contains remainders of the saturation spikes and the rings of the diffraction pattern. However, it is possible to identify some bright pixels $0\farcs9$ due northeast of the star (PA$=42\degr$) that may belong to the eastern lobe of the jet. The position angle between this bright feature and the southwestern jet is $71\degr$, matches the expected orientation of the jet. The problem is that the line joining both features goes north of the peak emission of the central object. At this moment we can not be certain whether this northeast emission is part of the counterjet, and we refrain ourselves from further analyzing it.   

\subsection{The molecular environment surrounding DO\,Tauri} \label{sec:rings}
\subsubsection{Three sets of CO rings}

\begin{figure*}
\centering
\includegraphics[angle=0, scale=0.65]{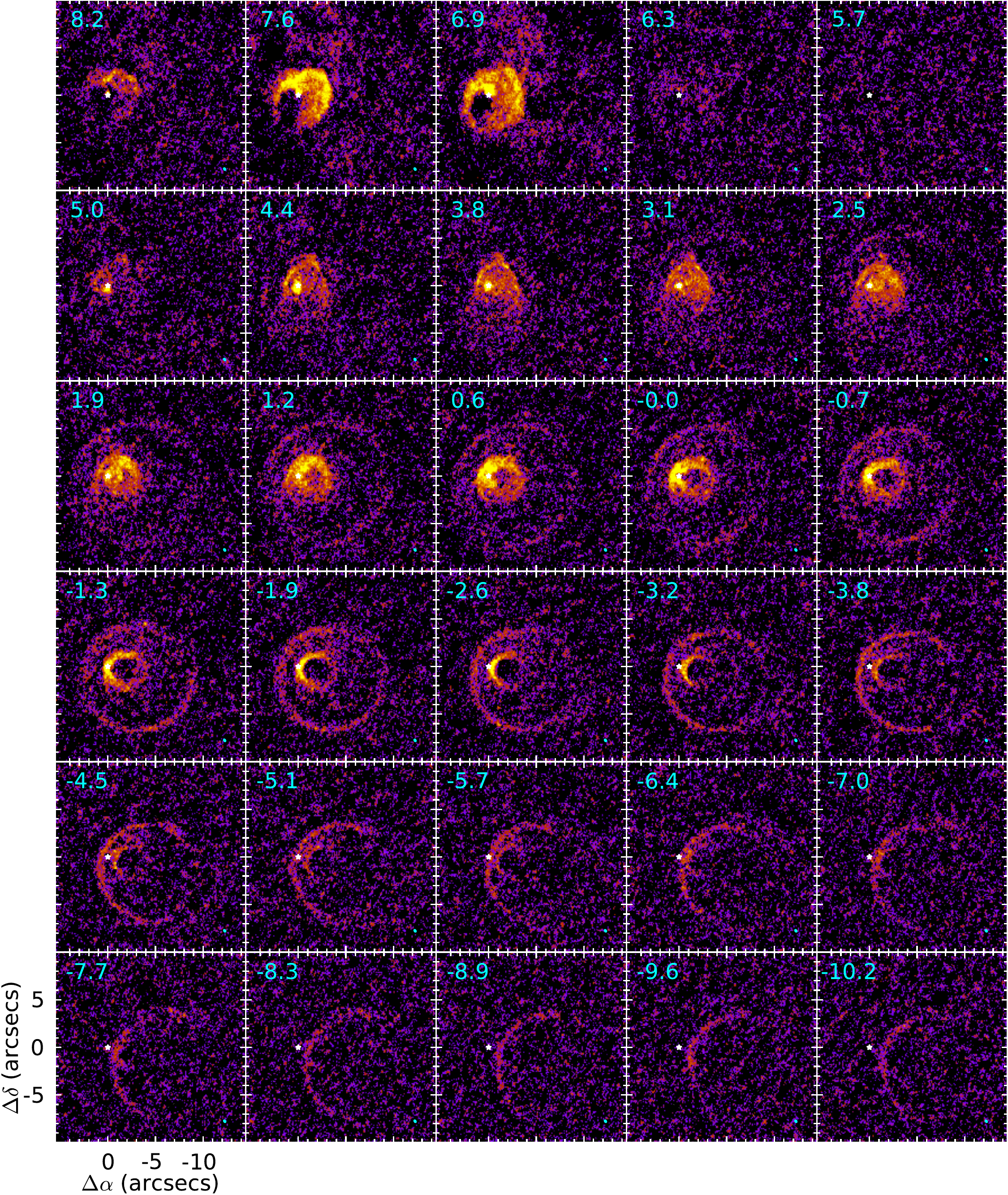}
\caption{Color scale image of the ALMA CO\,[J=2$\rightarrow$1] velocity channels from -10.2\kms to 8.2\kms. The color intensity range goes from 3\mjy to 30\mjy (2.1\,K to 21\,K). The position of the 1.3\,mm continuum peak is marked with a white star. The white dotted ellipse represents the fit to the large ellipse seen in the channel at -1.3\kms. The velocity is indicated in the upper left corner and the synthesized beam ($0\farcs21\times0\farcs16$; PA$=7\degr$) is represented as a cyan ellipse in the bottom right corner of each channel respectively.
 } 
\label{cube} 
\end{figure*}

Figure \ref{cube} shows the velocity cube of the \co emission surrounding DO\,Tauri from 8.2\kms to -10.2\kms. From -10.2\kms to -12.7\kms the emission is weak (see also figures in the Appendix) and at velocities larger than 8.2\kms, there is only emission from the disk. The cloud velocity is centered at 5.7\kms and the corresponding channel shows the typical interferometer emptiness due to the presence of extended emission and some absorption spatially coincident with the disk position (both are seen also in channels at 5.0\kms and 6.3\kms). The red-shifted emission spans $\sim 5$\kms while the blue-shifted emission spans $\sim18$\kms. This asymmetry in the CO line profile was already noticed by \cite{1995Koerner1} in a spectrum toward the young stellar disk. The CO emission surrounding the disk shows some elliptical structures of different sizes. To understand these CO structures, we split them in three sets of elliptical rings spanning different velocity ranges\footnote{Note however, that some of the rings appear incomplete, with a crescent-like structure. These rings show their northeast side brighter than their southwest side in a consistent manner with velocity.}.

\begin{figure*}
\centering
\includegraphics[angle=0, scale=0.65]{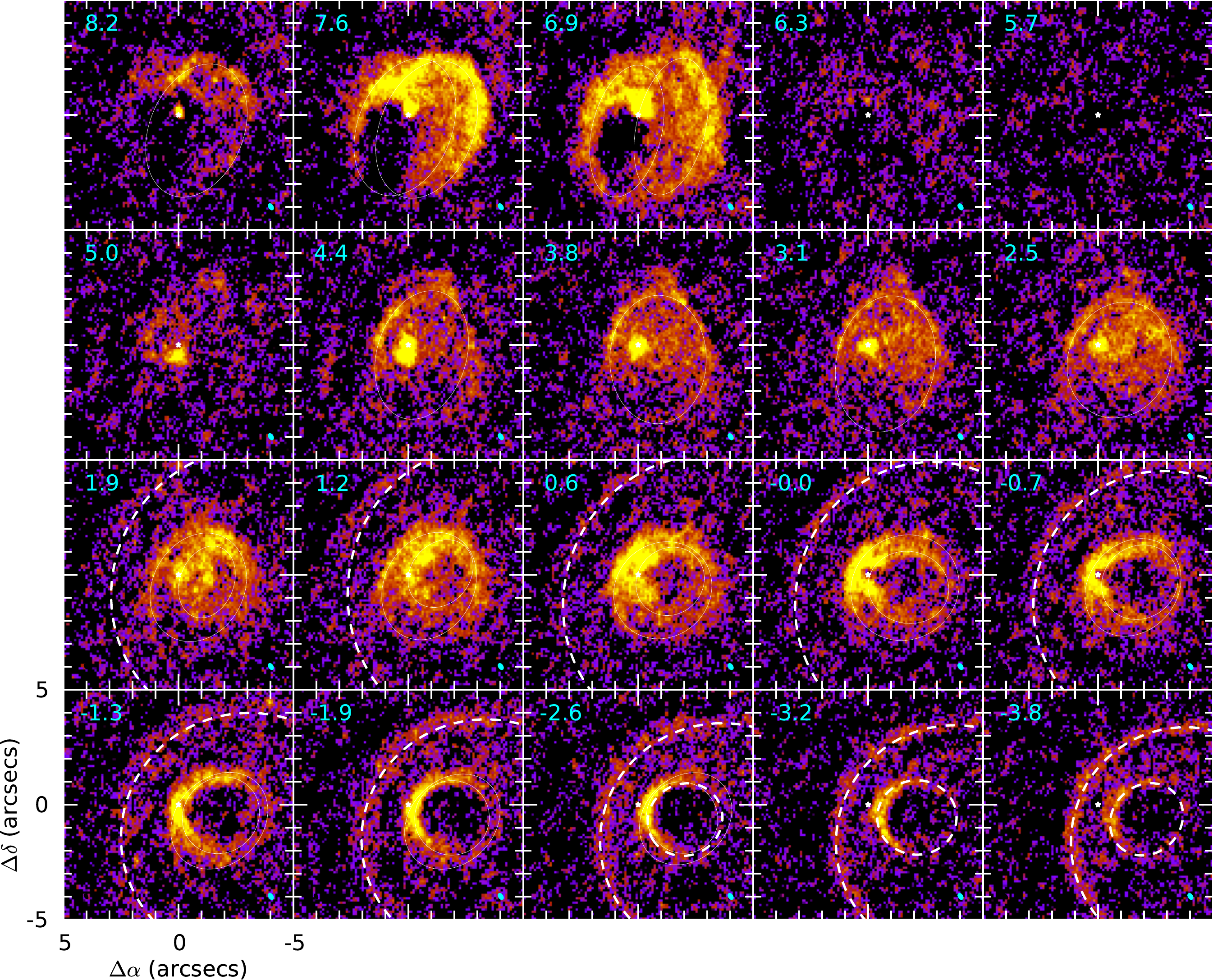}
\caption{\co emission of the central velocity channels demonstrating the manual ellipse fitting carried out. A star marks the position of DO\,Tauri disk's continuum peak, while the channel velocity and the synthesized beam are in the upper left and bottom right corners. Continuum/Dashed ellipses show manual/automatic fits carried out in the images.
} 
\label{elfit}  
\end{figure*}

A small scale set of 17 rings of average geometrical mean radius $1\farcs6$ (225\,au) and width between $0\farcs6$ and $1\farcs0$ (80-140\,au) can be identified in velocity channels from -8.3\kms to 1.9\kms. Here on we refer to these rings as set of Small Blue-shifted rings or SB rings. For instance, these rings are prominent in the -3.2\kms channel (the brightest CO structure at this velocity, Fig. \ref{elfit}), in which the disk of DO\,Tauri is outside the SB  ring, although it is close to its easternmost rim. The SB  rings are the brightest features seen in CO at blue-shifted velocities. Their northeast rim is brighter compared to their southwest rim, which fades (and is undetected) at more blue-shifted velocities 
(v$_{rad}<-2.6$\kms), making the rings incomplete.

In channels where  v$_{rad}>-2.6$\kms we identify a second set of 17 rings, which we name as Medium-size rings or M rings hereafter. These rings surround the smaller SB  rings from -2.6\kms to 1.9\kms, velocities at which both structures are present simultaneously (see for instance channels at 0.6, 1.2 and 1.9\kms in Fig.\ref{elfit}). Between 1.9\kms and 4.4\kms the M rings encompass the emission from the DO\,Tauri disk and are filled with some extended emission. At red-shifted velocities, two simultaneous (and equal) M rings explain the emission distribution morphology better (Fig. \ref{elfit}). These two rings spatially coincide with the northeast and northwest arcs observed by \cite{2008Itoh} using NIR and optical images. The M rings have an average geometrical mean radius of $2\farcs3$ (320\,au) and a width ranging between $0\farcs6$ and $1.1\farcs$ (80-150\,au). In general M rings are oriented northwest-southeast and show brighter emission toward their northern rim.

Figure \ref{cube} also show a third set of rings that encompass the M and SB sets of rings at larger scales, from -10.8\kms to 1.9\kms. We name them set of Large Blue-shifted rings (LB rings). The 23 LB rings have an average geometrical mean radius of $5\farcs7$ (about 800\,au) and a width between $0\farcs8$ and $1\farcs6$ (110-220\,au). They show stronger emission and are seen as complete rings between 1.2\kms and -4.5\kms. At more blue-shifted velocities they fade and only their northeast rims are detected.

\subsubsection{Ring fitting procedure} \label{sec:fits}
In order to make a more comprehensive analysis of the three sets of rings present in the data we fit 2D-ellipses to the emission of each ring channel by channel. We use an automatic fit (see below) only for those rings with a clear morphology, not blended or confused with other structures. For those rings with emission mixed up or complicated for our automatic procedure, a manual fitting was carried out trying to follow the brightest ridge of the rings. The results of such fits are summarized in Table \ref{trings} and can be seen in Figure \ref{elfit} and in the Appendix, in Figures \ref{large_red} and \ref{large_blue}. Automatic and manual fits are distinguished by a letter code in the mentioned table and by dashed or solid lines in the figures. 

For the automatic fits we used a Perl-PDL script with the Minuit\footnote{http://pdl.perl.org/?page=credits} tool \citep{1975James} which maximizes the product between the channel image and the equation of an elliptical ring with a negligible width compared with the synthesized beam size (we do not attempt to fit the width of the rings because it is variable along a single one). In order to determine proper uncertainties we used a normalization factor proportional to the rms of the velocity cube and to the length of each ring. Hence, Minuit derives an error for each parameter (coordinates center, major and minor axis of the ring and position angle) as the value that makes the fit deviating one times the value of the standard deviation of the merit function evaluated in noise regions. This procedure gives positional and size errors typically of the order of half the size of the synthesized beam ($\sim0\farcs1$). These automatic fits were not possible in channels with more complicated emission distributions. In those channels we perform manual fits to the emission trying to keep the fits as simple as possible (i.e., minimizing the number of rings). We estimate the uncertainties on the manual fitting empirically, by bringing the fits to the limit where the fitting ellipse escape the boundaries of the emission ring. In general, these errors are no larger than $\pm0\farcs3$ for the central coordinates, $\pm0\farcs3$ for the major and minor axes and $\pm15\degr$ for the position angle of the ellipses. Finally, as a conservative estimate we added $0\farcs1$ to the errors in position and sizes and $5\degr$ to the PAs. Furthermore, since some of the derived PAs have large uncertainties we decided not to quote the PA of the rings when their eccentricity $\sqrt{1-(Bmin/Bmaj)^2}$ is less than 0.35 (i.e., we consider these rings are roughly circular).

\subsubsection{Ring properties as a function of radial velocity}
\begin{figure*}
\centering
\includegraphics[angle=0, scale=0.65]{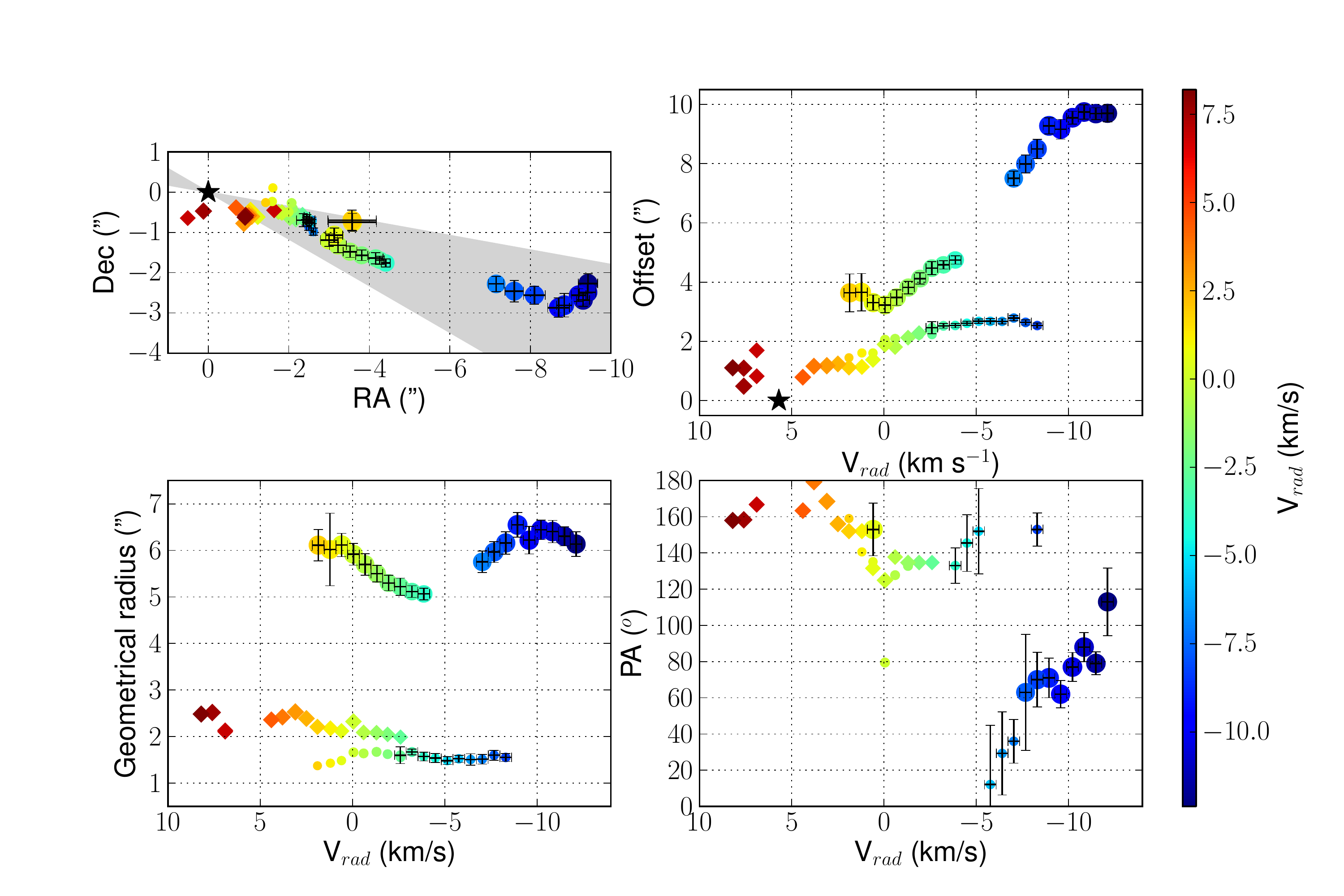}
\caption{\textbf{Top left:} Position of the center of the fitted ellipses with respect to the disk peak continuum emission. The points are colored depending on their radial velocity. Symbol sizes increase with ring sizes. M rings are represented by diamonds and SB and MB rings by circles. The position of the continuum peak of DO\,Tauri's disk is marked with a star. An overall linear trend can be seen spatially coincident with the direction of the optical jet (and its uncertainty), which is outlined by the shadowed area. \textbf{Top right:} Distance of the ring centers to the location of DO\,Tauri's disk as a function of their radial velocity. Trends of increasing velocity with distance from DO\,Tauri can be distinguished for different sets of rings. A star marks DO\,Tauri's position and the velocity of the cloud surrounding it (5.7\kms). \textbf{Bottom panels:} Geometrical mean radius and PAs of the ellipses as a function of their radial velocities.
} 
\label{analysis4} 
\end{figure*}

The ring fits show some general characteristic trends (some of them as a function of radial velocity) that are shown in Figure \ref{analysis4}. First, let us notice that the centers of the elliptical rings with respect to the position of the disk peak continuum emission (we will refer to these as centers or ring centers from now on for simplicity) are linearly aligned with the position of DO\,Tauri (top left panel in Fig. \ref{analysis4}). A linear fit to the ring centers (taking only those rings with radial velocity lower than 5.7\kms) gives a position angle of $253.6\degr\pm0.1$ and an alignment better than $0\farcs1$ with the DO\,Tauri's disk position. This position angle matches the orientation of the optical jet reported by \cite{1997Hirth}. In addition to this trend, the centers of the SB  and M rings are closer to the young stellar disk than those of the LB  rings.

Second, the positions of the ring centers are also correlated with their corresponding radial velocities. The top right panel in Fig. \ref{analysis4} shows that the larger the distance between the ring center and the young star, the bluer (the larger, or more negative) is its velocity. Moreover, while a linear trend makes a reasonable fit for the LB data, the trends for the SB and M rings are not just linear correspondences. A closer look into these data shows that the velocity increases roughly exponentially with distance from the source, although more points (higher spectral resolution) would be needed to better sample these trends. For now we can confidently say that the radial velocity increases with the distance from the young star and the increment is more steep further away from the star.

In third place, the bottom left panel in Fig. \ref{analysis4} shows the geometrical mean radius of the rings as a function of the radial velocity. The three sets of rings are clearly segregated from each other. SB  rings are below $1\farcs9$ (260\,au), M rings are between $1\farcs9$ and $3\arcsec$ (560\,au) and LB  rings are larger than $4\arcsec$. Despite the size of the rings within each set does not change much, the M rings slightly decrease their size monotonically toward bluer velocities and the LB  rings monotonically decrease their radii from $6\arcsec$ to $4\arcsec$ and then increase them again monotonically up to more than $6\arcsec$.

The fourth plot (bottom right panel) in Figure \ref{analysis4} shows the change on the ring orientations as a function of radial velocity. Most of the PAs of the SB  rings are in a range of 40$\degr$ around PA$=140\degr$ (note that a quarter of these rings are quite circular and just a few of them are closer to PA$=25\degr$), while those of the M rings seem to linearly vary from $120\degr$ (blue-shifted velocities) to $180\degr$. The orientation of the LB  rings is more uncertain since these rings are quite round; in any case their PAs do not change more than 20$\degr$ overall.      

\begin{figure*}[!htb]
\minipage{0.48\textwidth}
  \includegraphics[width=\linewidth]{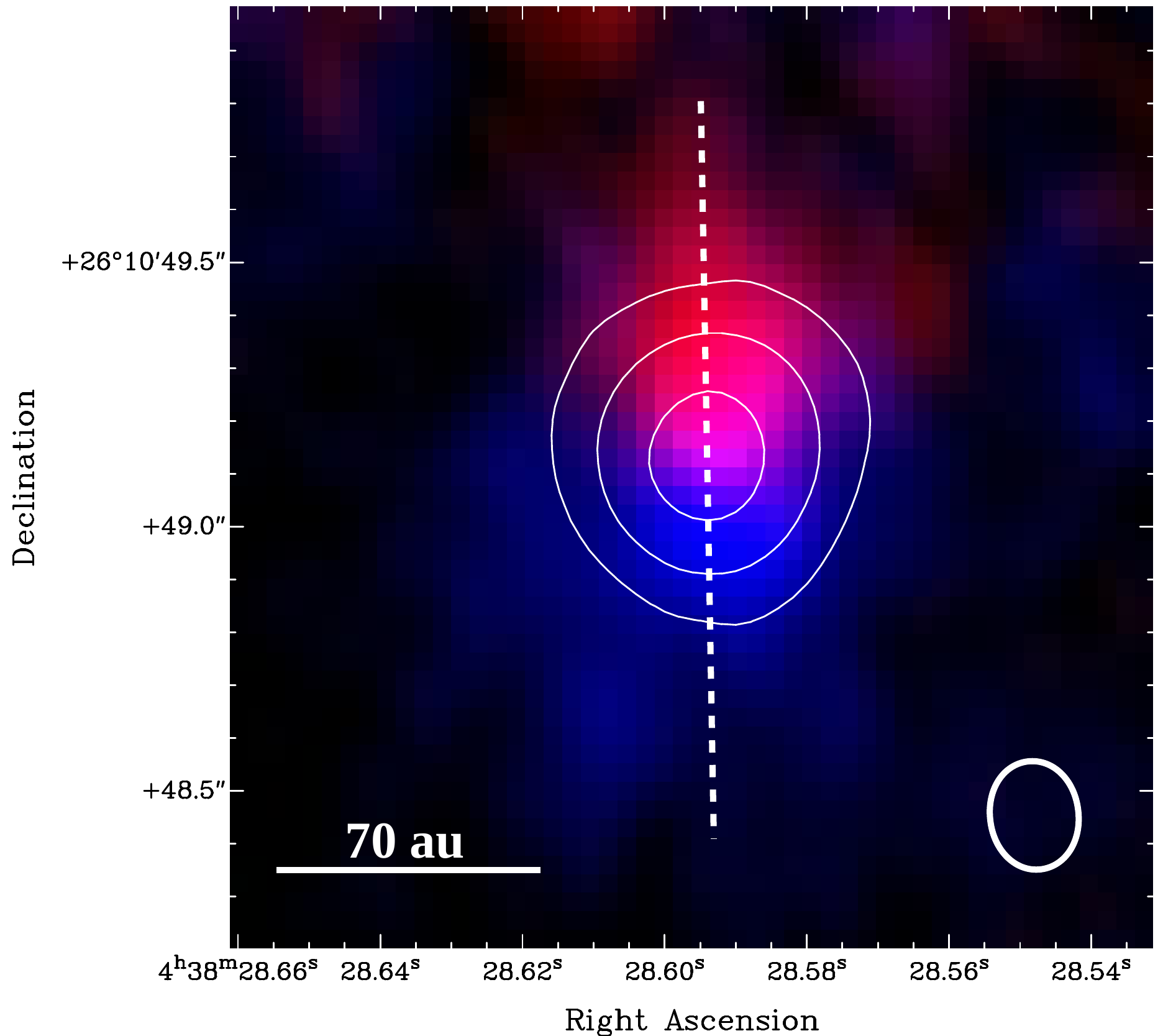}
\endminipage\hfill
\minipage{0.48\textwidth}
  \includegraphics[width=\linewidth]{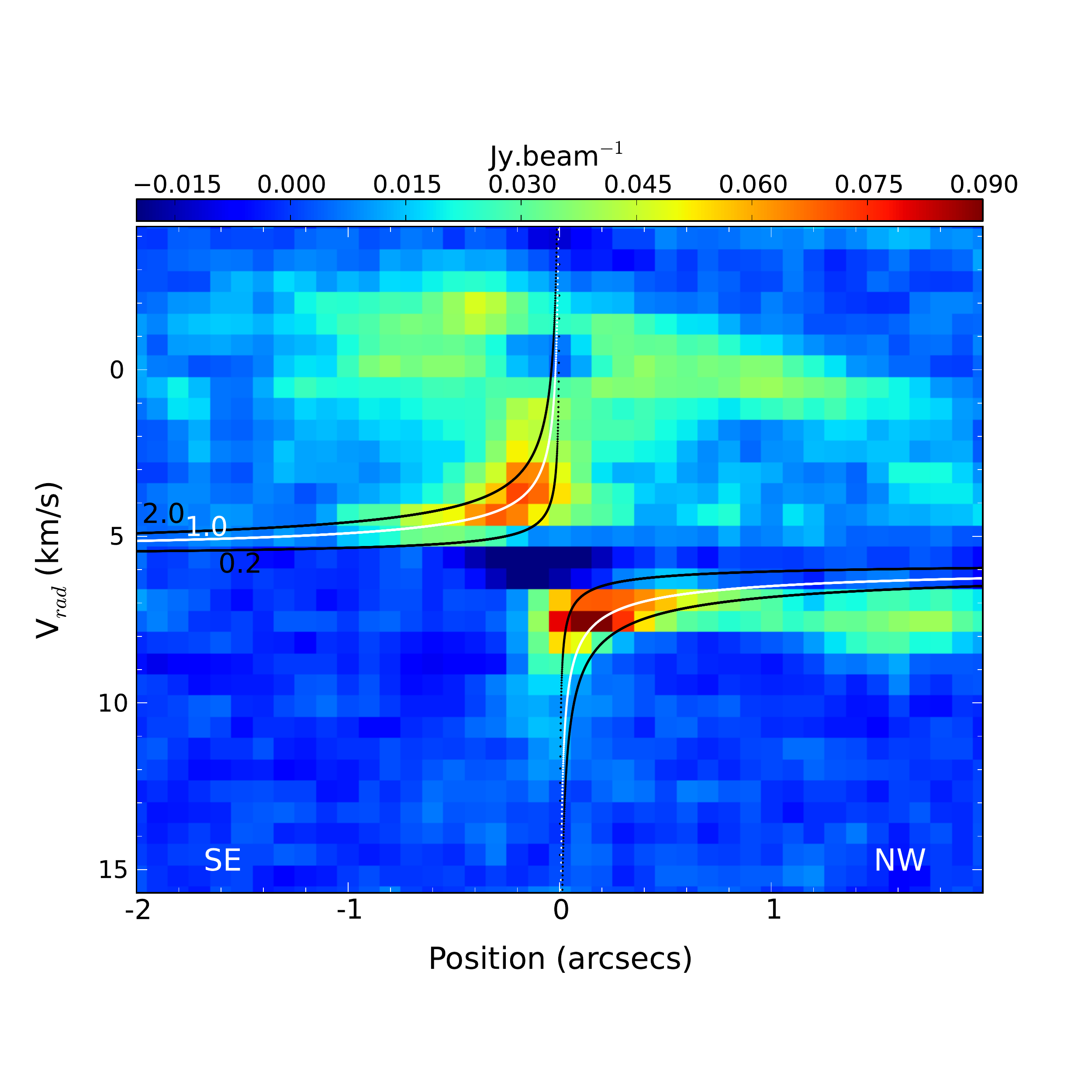}
\endminipage\hfill
  \caption{\textbf{Left:} The image shows the ALMA CO\,[J=2$\rightarrow$1] integrated blue-shifted [0.6,5.0]\kms and red-shifted [6.9,12.7]\kms intensities (color-scale) overlaid with the 1.3\,mm high-angular resolution continuum emission (white contours) toward the inner region surrounding DO\,Tauri. 
The 1.3\,mm continuum emission is represented by contours at $30\%$, $50\%$ and $80\%$ of the peak emission (24.546\mjy).
   \textbf{Right:} Position-velocity diagram of the CO\,[J=2$\rightarrow$1] emission along a cut passing through the peak of the disk emission and a PA of $179\degr$. The cloud velocity is at 5.7\kms. Curves show Keplerian velocity patterns of an inclined disk (i$=19\degr$) with central masses of 0.2, 1.0 and 2.0\msun. At blue-shifted velocities the presence of extended emission complicates identification of the disk's kinematics.}
  \label{fdisk}
\end{figure*}

\section{Analysis and discussion} \label{sec:dis}
In this section we try to interpret the ALMA \co observations toward DO\,Tauri. Although more information is needed to fully understand the origin of these molecular structures, we discuss the possibility of an outflow entrained by an episodic wide-angle wind or jet from the young star. 
The usual appearance of protostellar outflows is that of a biconical parabola, but the geometry of an outflow seen almost pole-on should reveal the cavity walls of the outflow as a ringed structure when imaged in a velocity cube, in a similar manner to the present CO observations from the surroundings of DO\,Tauri. This hypothesis have further support when taking into account the very inclined orientation (almost face-on) of DO\,Tauri's circumstellar disk (section \ref{sec:disk}) if, as expected, it is perpendicular to the jet/outflow direction. 

The interaction between the interstellar medium and DO\,Tauri itself, which may be moving supersonically in the molecular cloud after a close encounter with HV\,Tauri in the past \citep{2018Winter} could somehow account for some of the observed features, but we underweight this possibility as DO\,Tauri seems to be roughly co-moving with the Taurus molecular cloud (the cloud velocity coincides with the kinematic center of the DO\,Tauri disk, Fig. \ref{fdisk}), and its GAIA\,DR2 proper motions are similar to those of the stars from its vicinity.\footnote{For DO\,Tauri, Gaia\,DR2 proper motions are $\mu^*_{\alpha}=6.1\pm0.1$\,mas\,yr$^{-1}$ and $\mu_{\delta}=-21.34\pm0.09$\,mas\,yr$^{-1}$; the average proper motions of the objects observed by Gaia\,DR2 toward a region inside a 40\,arcminutes radius centered at DO\,Tauri, with parallaxes ranging between 6.2\,mas and 8.3\,mas ($\pm20$\,pc from DO\,Tauri's distance of $\approx$140\,pc) and with parallax errors of less than 0.25\,mas are $\mu^*_{\alpha}=6\pm1$\,mas\,yr$^{-1}$ and $\mu_{\delta}=-21\pm2$\,mas\,yr$^{-1}$ (note that we removed a few outliers with large proper motions for this estimate); this results in DO\,Tauri having a tangential velocity with respect to these objects of 1$\pm$3\kms (PA$=162\degr\pm120\degr$).} 

In this section we add more support to the outflow hypothesis, extending it to analyze the correlations between the velocity and the physical characteristics of the rings, estimate the dynamical age of the outflow shells traced by the rings, as well as their mass and energetics.

\subsection{An outflow from DO\,Tauri}
CO is known to be a good outflow tracer in star-formation regions, although it can be also encountered as part of the disk, the envelope and the parental cloud. In DO\,Tauri, however, the \co emission is both spatially extended (spreading much further than the size of the disk) and extended in velocity (over about 20\kms), so it seems at least plausible that the SB , M and LB  rings seen in the vicinity of DO\,Tauri are part of the material dragged by the high-velocity jet reported in the literature \citep{1994Hirth} and detected in the HST images presented here. Indeed, the sets of elliptical rings seen in CO have their centers aligned with the position of DO\,Tauri (with the smaller rings closer to the young star than the larger ones) and the PA of the optical jet ($260\degr\pm1\degr$) roughly coincides the PA of the aligned ring centers ($253.6\degr\pm0.1$). 

Usually, outflows show a parabolic biconical shape because they are seen close to the plane of the sky, but in the case of DO\,Tauri we do not see this parabola. We interpret the three families of elliptical rings as three different shells of entrained material seen at different radial velocities. In this scenario we see the outflow from an almost pole-on perspective. Indeed, if the jet and outflow are launched perpendicular to the circumstellar disk plane, the inclination of the jet with respect to the line of sight would be the same as that of the disk, $i=19\degr$. From now on we will assume this is the case and we use this inclination to correct for the estimated physical properties when adequate. Taking this inclination into account, the dimensions of the outflow would be 4070\,au$\times$1960\,au from the young star to the further ring detected by the ALMA observations. This implies an opening angle of $27\degr$ (a collimation factor of 2.1). The size of the CO structure, its opening angle and the radial velocity extent, all agree with the usual measures of CO outflows from protostars and young stars \citep[from a few thousands au to a few pc in length, opening angles typically between $20\degr$ and $60\degr$ and radial velocity spreads between 10\kms and 100\kms;][]{2006Arce,2016Bally}. When compared with one of the most recent paradigmatic outflows from Class\,0 objects such as HH\,46/47 \citep[about 60000\,au long and more than 10000\,au in width][]{2013Arce}, this one would be a small (young?) outflow, but in the event we are only detecting the inner shells of the DO\,Tauri outflow then the sizes match better, being the outflow from DO\,Tauri still more collimated, for HH\,46/47 inner shells fit in a box of about 5000\,au side \citep{2016Zhang,2019Zhang}. If compared with another classical outflow from a Class I/II young star, DG\,Tau\,B \citep[see e.g.,][]{2015Zapata}, both have sizes of the same order of magnitude (DG\,Tau\,B's is 2550\,au long and about 1400\,au wide).
It is also significant that the outflow from DO\,Tauri would not have a clear red-shifted lobe. This is not a huge obstacle for the outflow hypothesis since other monopolar outflows (probably due to environmental inhomogeneities) have been reported in regions such as GGD\,27 or HH\,30 \citep[][]{2013FernandezLopez,2018Louvet}.

\begin{figure*}[!htb]
\minipage{0.50\textwidth}
  \includegraphics[width=\linewidth]{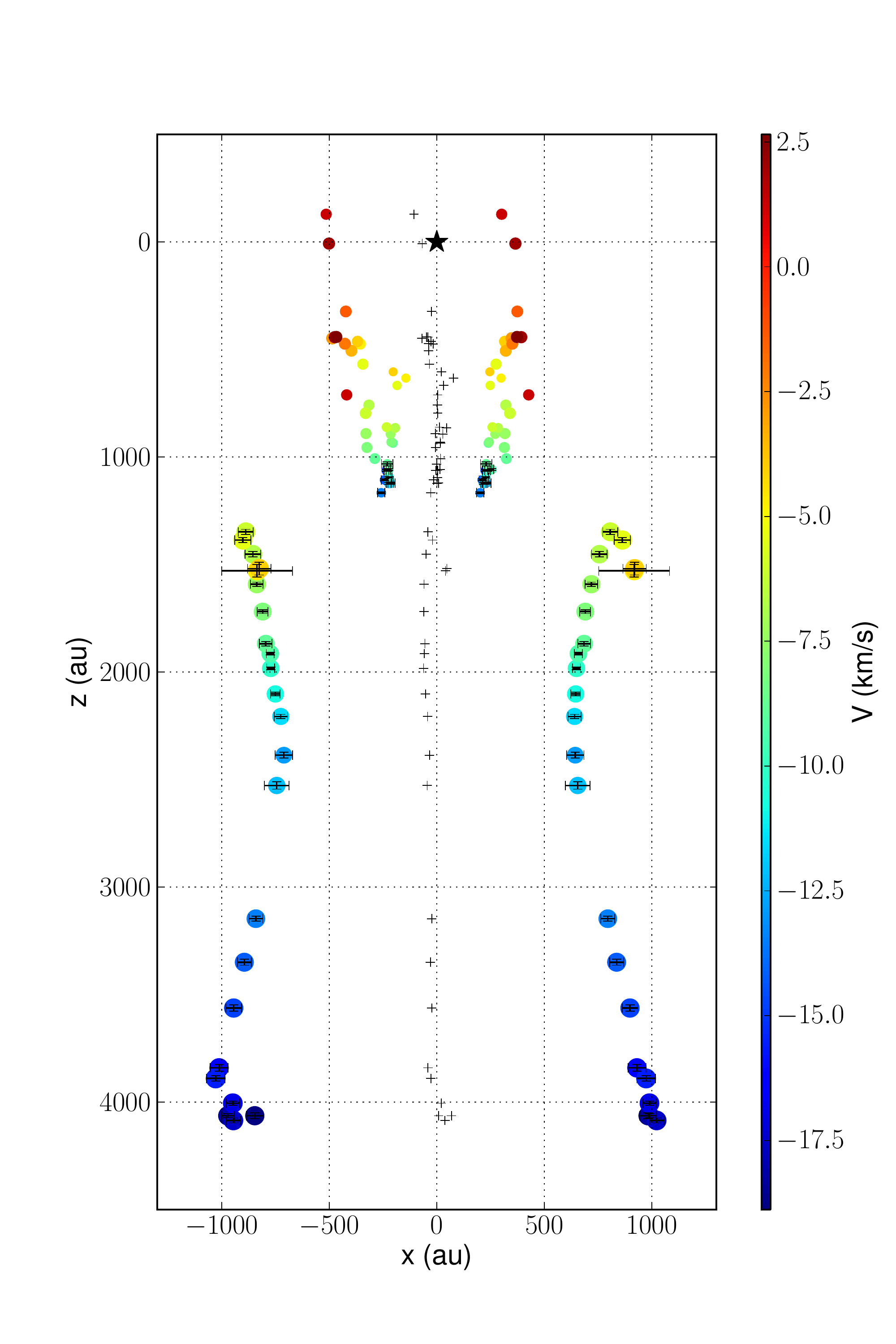}
\endminipage\hfill
\minipage{0.50\textwidth}
  \includegraphics[width=\linewidth]{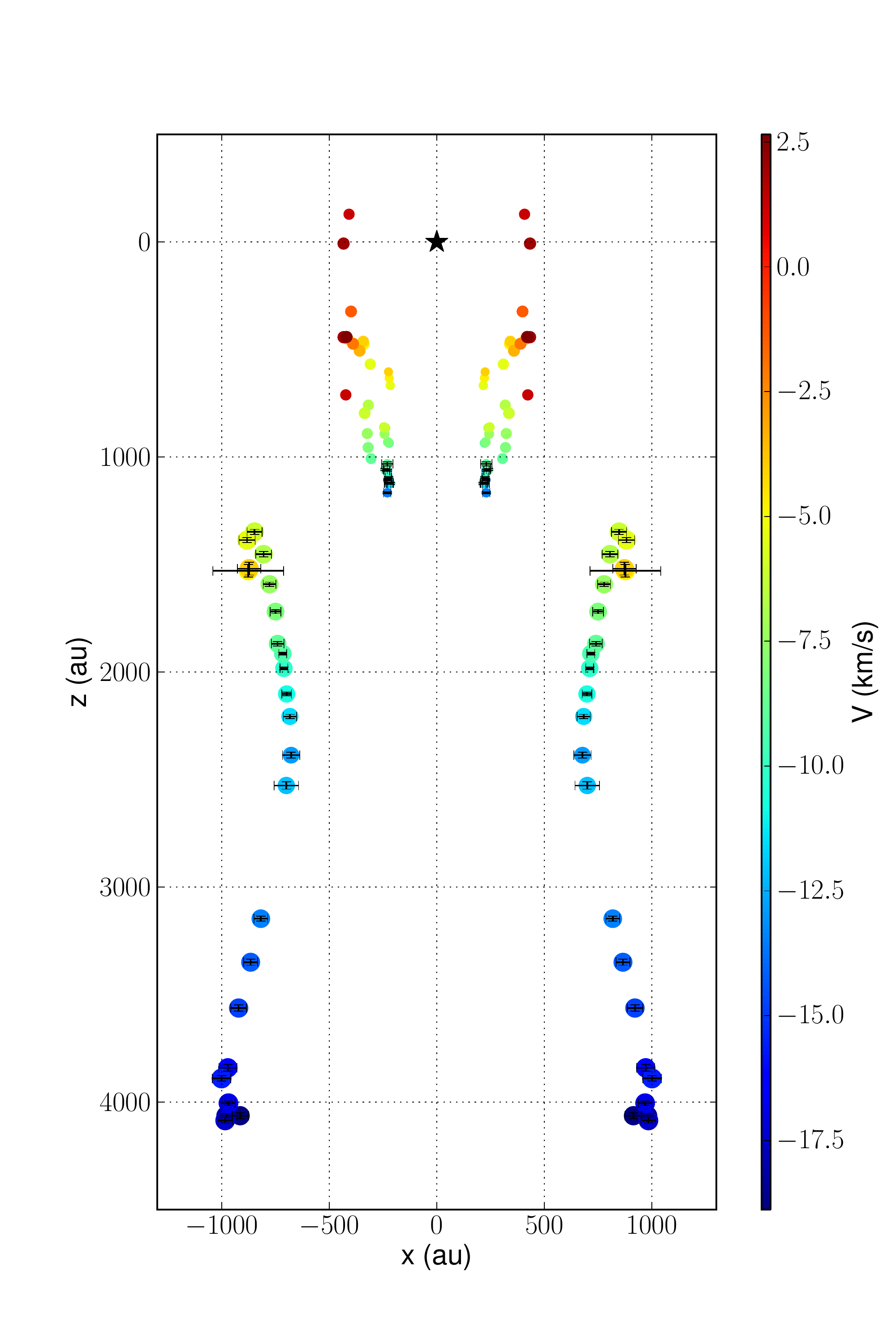}
\endminipage\hfill
  \caption{\textbf{Left:} Reconstruction of the outflow from DO\,Tauri from the ring fits (ring center and major axis), assuming that the rings are circunferences with radius equal to the measured semi-major axis and the outflow is $19\degr$ inclined with respect to the line of sight. The crosses mark the position of the ring centers after a $253.6\degr$ rotation in the plane of the sky, so that the blue lobe of the outflow matches the positive z-axis. Distances and velocities are corrected by the outflow inclination. A star marks the position of DO\,Tauri.
   \textbf{Right:} Same but forcing the center of each ring to be in a straight line along a putative linear outflow axis coincident with the z axis.}
  \label{foutflow}
\end{figure*}

Finally, let us show a reconstruction of a spatial cut parallel to the outflow major axis based on this scenario (Fig. \ref{foutflow}). First, we rotate the (RA, Dec) coordinates of the ring centers into the $(x,z\prime)$ plane, where the projected $z\prime$ axis is oriented with the outflow axis taking positive values southwest of the young star. We used an outflow PA of 253.6$\degr$ to rotate the coordinates. Second, we correct for the inclination of the outflow with respect to the line of sight, considering it is perpendicular to the disk, to get the unprojected coordinates along the $z$-axis. After this, we assume that the rings are perfect circumferences with radii equal to the semi-major axis of the fitted ellipses. That is how we build the left panel of Figure \ref{foutflow}. The morphology outlined by the points in this figure reminds the shape of protostellar outflows. In the left panel two cylinder-like structures are outlined: one with smaller radius composed of the SB  and M rings extending up to $\sim1200$\,au, and a second one with a larger radius that doubles this size reaching $\sim4000$\,au (LSB rings). The right panel of Figure \ref{foutflow} has a further step in the reconstruction, by forcing the position of the rings to lie in the $z$-axis (i.e., we force them to be at $x=0$). We remove this way any wiggling of the outflow or errors in measuring the ring center positions. The result is a symmetric outflow, but here we can notice additional interesting details. On one side, the inner part of the outflow ($<1200$\,au) comprises two concentric structures: one with a roughly constant width (mostly formed by the SB  rings) and a wider one whose width seems to decrease smoothly with the distance to the star (the M rings). In contrast, the widest LSB shell ($>1200$\,au) smoothly decreases its aperture upto 3000\,au from the star and from this point increases it again. 

\subsection{Velocity gradients and dynamical ages from different shells}\label{s_ages}

In Fig. \ref{analysis4} each set of rings shows a different velocity gradient (SB , M and LB). These velocity gradients show an increase in velocity with distance, sometimes steeper than a simple linear trend. 
Although the gradients are not linear they show a similar trend to the Hubble-law behavior found in the cavity walls of some molecular outflows \citep[see for instance][]{2016Zhang,2019Zhang}. Radial wide-angle winds can account for such Hubble-laws provided they have an angular dependent force (e.~g., $\propto 1/\sin^2(\theta)$, where $\theta$ is the polar angle and $\theta=0\degr/180\degr$ correspond to the direction perpendicular to the disk plane) and sweep-up gas in a medium with a specific density structure (e.~g., $\propto\sin^2(\theta)/r^2$). Under this model the resulting outflow walls have a roughly parabolic morphology that expands radially with a constant speed, independent of the outflow shell size. The velocity V$(\theta)$ is different for each direction $\theta$, but constant in time. As a result, at any given age t$_{dyn}$, the shell shows an apparent Hubble-law, with patches of the shell closer to the source being slower than further ones with V$(\theta)=$R$(\theta)/$t$_{dyn}$\citep{1991Shu,1996Li,2000Shu,2000Lee}.

\begin{figure}
\centering
\includegraphics[angle=0, scale=0.48]{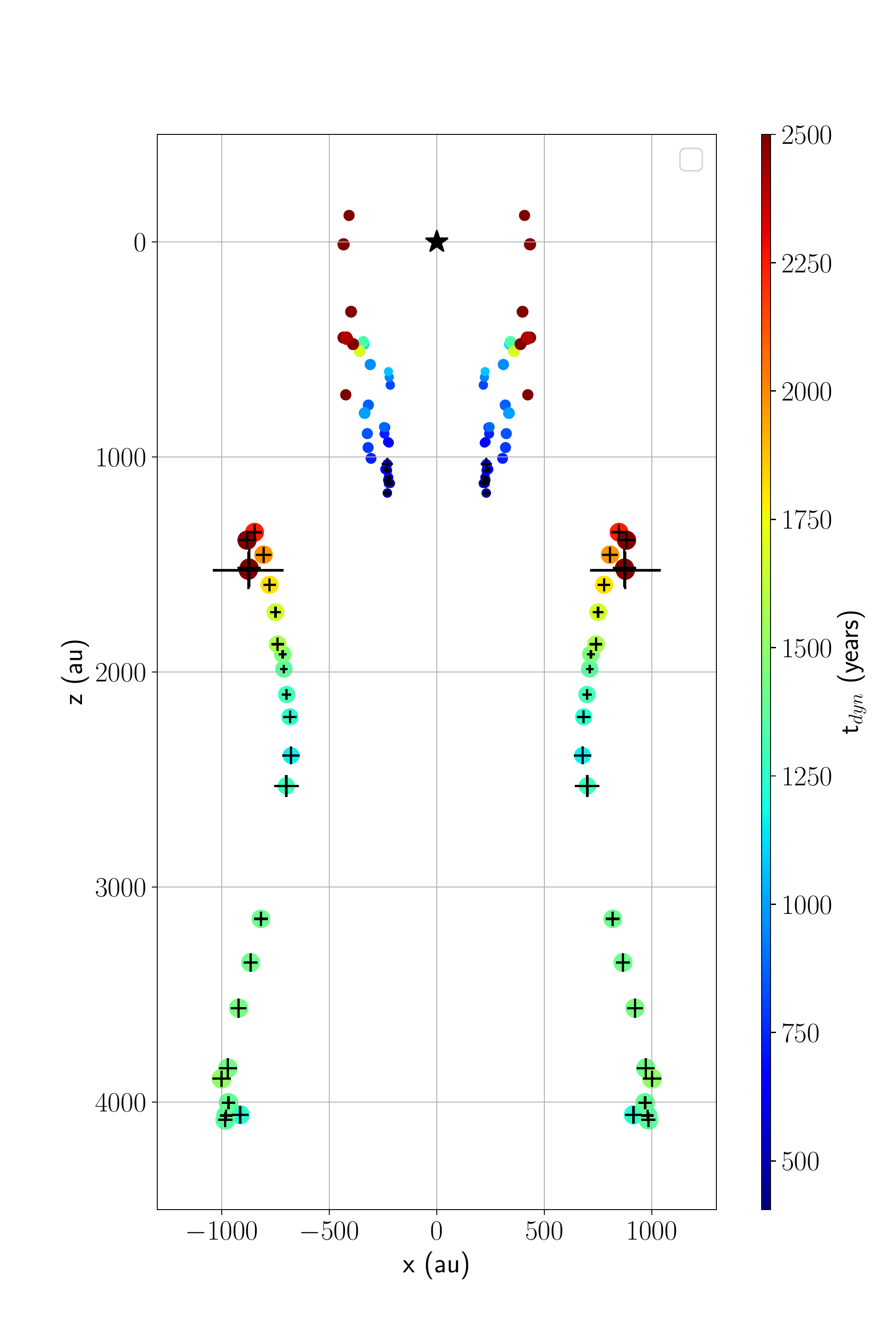}
\caption{Same as in right panel of Fig. \ref{foutflow} but using the dynamical ages of the rings as the color code.
} 
\label{fages2}  
\end{figure}

\begin{figure*}
\includegraphics[angle=0, scale=0.4]{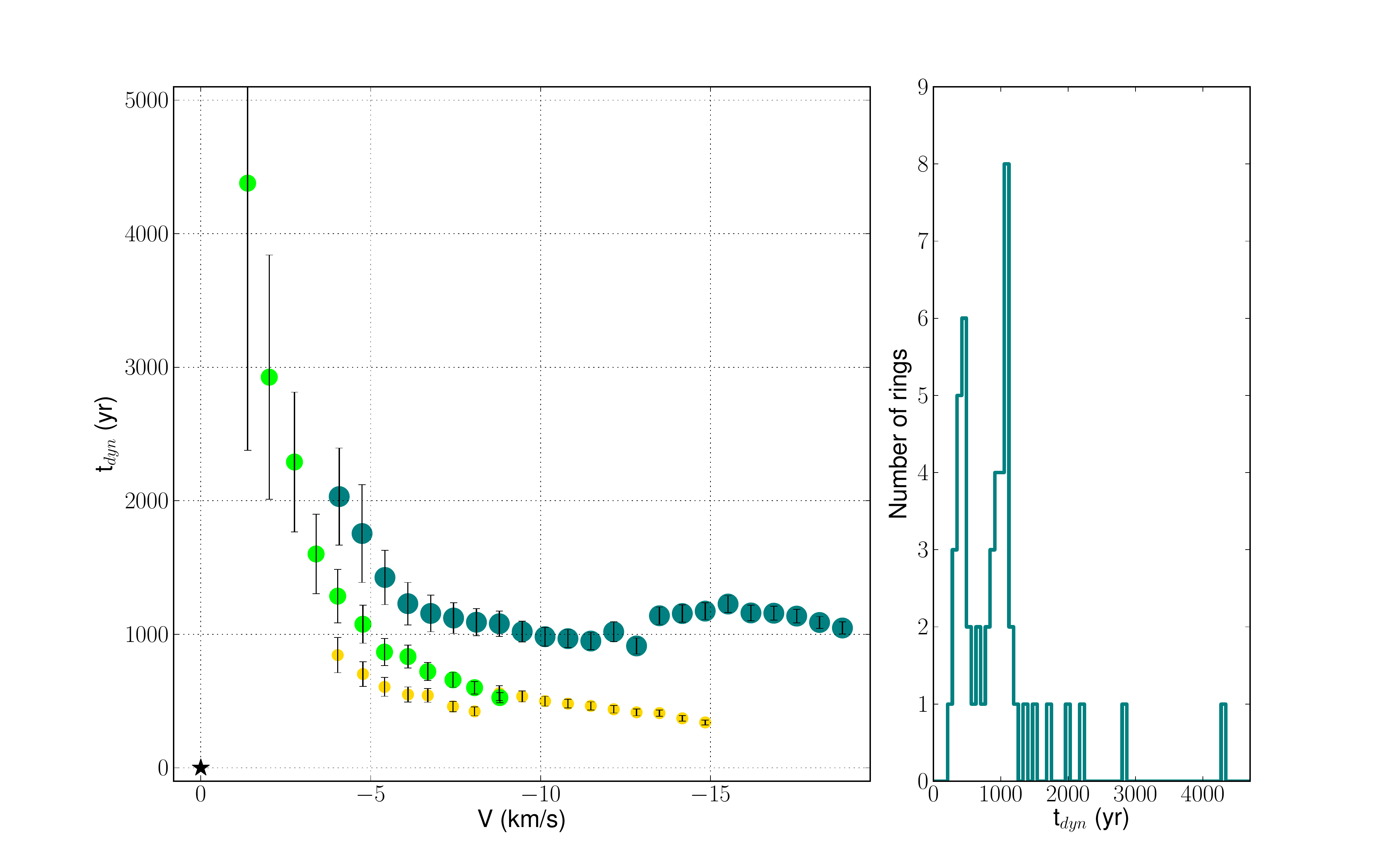}
\caption{\textbf{Left panel:} dynamical ages of the molecular rings as a function of the velocity of the blue-shifted lobe of the outflow once corrected for the inclination. Each color and symbol size represent a different outflow shell: large blue, medium-sized green and  small yellow circles correspond to LB, M and SB shells, respectively. \textbf{Right panel:} histogram of the dynamical ages of the outflow rings.    
 } 
\label{fages} 
\end{figure*}

Assuming the wide-angle wind model we estimate the dynamical age of each shell by computing the dynamical time of each ring composing it as t$_{dyn}$=R/V, being R the deprojected distance from the young star to the edge of a ring ($R = \sqrt{z^2+B_{maj}^2}$, where z is the distance from the star to the center of the ring and B$_{maj}$ is the ring's major axis) and V being an averaged value of the deprojected velocity of the molecular gas of the rings, respectively. We take $V\sim V_{radial}/\cos{i}$ ($i=19\degr$ is the inclination of the outflow) as an averaged value of the deprojected velocity of the gas in an individual ring\footnote{Note that, assuming that the shell expands purely in a radial direction, the deprojected velocity of the gas in a ring varies from $V_{radial}/\cos{i+\epsilon}$ and $V_{radial}/\cos{i-\epsilon}$, where $\epsilon$ is the semi-opening angle of the ring when viewed from the source.}. Figure \ref{fages2} shows in a graphical way the spatial distribution of dynamical times along each shell surface in the reconstructed sketch of the outflow (see also Fig. \ref{foutflow}). There is a notorious dynamical age segregation between the SB and the LB outflow shells (see right panel on Fig. \ref{fages}), indicating that there are at least two different events of outflow entrainment \citep{2019Zhang}. We also consider the M rings as a third outflow shell that partly encompasses the SB shell.

While the dynamical ages of the rings composing the LB and SB shells are mostly distributed in a 300\,yr lapse, the dynamical ages of the M rings are quite spread (due to low velocities and relatively larger sizes). Among them, the more well-delineated M rings between -1.9\kms and 1.2\kms have dynamical ages ranging 600\,yr to 1100\,yr (700$\pm100$\,yr on average). These M rings comprise a shell that surrounds part of the the SB outflow shell, which appears $\sim300$\,yr younger. In contrast, the M rings closer to the cloud velocity and those that are red-shifted show extended emission inside\footnote{Whether this extended emission is part of the most inner part of the outflow or is part of the remains of a circumstellar envelope can not be decided with the present data.} and have dynamical ages from 600\,yr to 5000\,yr, with deprojected gas-velocities as low as 1.3\kms respect to the cloud velocity. These rings subtend the largest angles from DO\,Tauri in projection. 

Within the individual shells, not all the rings have the same dynamical time, unlike one would expect in the wind-driven model \citep{2000Lee}. These differences can be accounted for if there are any inhomogeneities in the ambient cloud that may decelerate the outflow material at different rates in different directions $\theta$. Moreover, deviations from the model are also expected if the outflow is not steady \citep{2001Lee} as is the case for DO\,Tauri, where multiple shells are observed.

An interesting alternative scenario for the DO\,Tauri outflowing shells is that they are generated by successive jet-driven bowshocks \citep{2001Ostriker}. In this model, a narrow jet produces an expanding bowshock when bumping into the ambient medium. The velocity of the outflow thus created increases as a power-law with increasing distance to the source. This model implies shorter dynamical times at larger distances from the source, which is the trend observed along the LB and SB shells (Fig. \ref{fages2}). The rings with the largest dynamical times in these shells show the largest opening angles too, and have gas with lower projected velocities (Fig. \ref{foutflow}).

Finally, from left panel of Figure \ref{fages} it can be inferred that within each shell, the dynamical times increase very steeply at lower velocities. This strongly suggests a slowing down of the shells close to the star where a more dense ambient material is expected. In particular, it should be noted that due to possible outflow deceleration, the measured dynamical times are upper limits to the true age. Hence, the best estimate of each shell age is obtained by measuring the dynamical time at large distances, where $t_{dyn}$ is almost constant and the deceleration does not seem to be too severe. We therefore estimate the dynamical age of the observed outflow in 1090$\pm$90\,yr, considering the dynamical time of the furthermost LB shell. The mean averaged dynamical times of the three shells are gathered in Table \ref{tglobal} (error bars for these values are derived from rms times).

\subsection{CO masses, momentum and energy of the rings}
Measuring gas masses using \co has several problems \citep{2001Arce,2014Dunham,2016Zhang} because this molecule is sometimes contaminated with emission from the disk and the natal cloud and can be optically thick. Here, we make an attempt to estimate the mass, linear momentum and kinetic energy of the molecular ringed structure found in DO\,Tauri, to compare these quantities with those of young stellar outflows.

Since we do not have at hand observations of any optically thick tracer we assume an excitation temperature of 25\,K as a proxy for the mass, momentum and energy estimates. Similar values have been used for outflows of other Class\,II young stars \citep[usually T$_{ex}<30$\,K, see e.g. CB\,26 and HH\,30,][]{2009Launhardt,2018Louvet}. Using this T$_{ex}$ we found that some channels are optically thick at the peak emission. This choice sets a lower limit to the mass derived from now on.

Despite the existence of \co optically thick emission at the brightest parts of the rings in several velocity channel (mainly in the SB  and M rings), large areas have brightness temperatures between 3\,K and 8\,K which translate into optical depths ranging between 0.2 and 0.5. Therefore we can just make a crude estimate of the mass, momentum and energy of the outflow by assuming local thermodynamic equilibrium and optically thin CO emission. 
Note in addition that it is also possible that the ALMA observations are filtering out part of the extended material of the outflow. With these caveats in mind, we estimate first column densities and from the shell masses, momenta and kinetic energy, using an analogue procedure to that in \cite{2008Pineda} and \cite{2016Zhang}, with expressions adapted to the \co transition following \cite{2015Mangum}. 
In particular, we use the following equation to derive the column density for the optically thin case: 
\begin{equation*}
\begin{split} 
 N_{tot}(CO)=\frac{3h}{8\pi^3\mu^2S}\cdot\frac{Q_{rot}}{g}\frac{~e^{E_u/kT_{ex}}}{e^{h\nu/kT_{ex}}-1} \\
 \cdot\frac{\int{T_B}dv}{[J_{\nu}(T_{ex})-J_{\nu}(T_{bg})]}=\\ 
 =\frac{3h}{8\pi^3\mu^2J}\cdot\frac{Q_{rot}~e^{E_u/kT_{ex}}}{e^{h\nu/kT_{ex}}-1}\cdot\frac{\int{T_B}dv}{[J_{\nu}(T_{ex})-J_{\nu}(T_{bg})]}
\end{split}
\end{equation*}
which for the particular transition \co can be expressed as
\begin{equation*}
\begin{split}
 N_{tot}(CO)=\frac{1.196\times10^{14}~(T_{ex}+0.921)~e^{16.596/T_{ex}}}{e^{11.065/T_{ex}}-1}\\
 \cdot\frac{T_B\Delta v}{[J_{\nu}(T_{ex})-J_{\nu}(T_{bg})]}.
\end{split}
\end{equation*}
To obtain the latter expression we used the line strength $S=J/(2J+1)=2/5$, the dipole moment $\mu=1.1011\times10^{-19}$\,StatC\,cm, the partition function $Q_{rot}=kT_{ex}/(hB_0)+1/3$ with a rigid rotor rotation constant $B_0=57.635968$\,GHz, the degeneracy $g=2J+1=5$, the T$_B$ in K and the $\Delta v$ velocity interval in \kms.     
From this, we estimate the mass by doing $M=\mu m_h \Omega N_{tot}/X_{CO}$, using a mean molecular weight $\mu=2.8$ times the atomic hydrogen mass and a CO abundance of $X_{CO}=10^{-4}$ ($\Omega$ is the area of the rings).
We further correct the momenta and kinetic energy by the adopted outflow inclination (we use $i=19\degr$ in $v_{outflow}=v_{rad}/\cos{(i)}$, where $v_{rad}$ is the line-of-sight or radial velocity). The results for each of the three outflow shells are summarized in Table \ref{tglobal}, considering the M shell as the part with $V<-4$\kms, thus excluding the red-shifted emission and the emission close to the cloud velocity contaminated by the DO\,Tauri circumstellar disk. The LB shell has approximately twice the mass of the SB shells. The three shells carry out a comparable amount of energy and momentum. Globally, the lower limits for the total mass of the blue-shifted outflow, the total linear momentum and the total kinetic energy are $1.3\times10^{-4}$\msun, $1.1\times10^{-3}$\msun\kms and $1.1\times10^{41}$\,erg. 

Moreover, with the mass and the dynamical age of the blue-shifted lobe of the outflow we can now derive an estimate (lower limits due to possible optically thick CO emission, T$_{ex}$ uncertainty and t$_{dyn}$ being an upper limit of the outflow age) of its average mass load rate ($1.1\times10^{-7}$\msun\,yr$^{-1}$) and its average rate of linear momentum injected by the wind/jet into the CO outflow ($1.0\times10^{-6}$\msun\kms\,yr$^{-1}$). Table \ref{tglobal} contains these quantities estimated for the three individual shells. There is a striking good agreement of the $\dot{M}$ and the $\dot{P}$ among the three shells, with average values $\dot{M}_{shells}=5.5\times10^{-8}$\msun\,yr$^{-1}$ and $\dot{P}_{shells}=4.7\times10^{-7}$\msun\kms\,yr$^{-1}$. These values may be compared with the mass-loss rate and its associated momentum rate extracted from the optical observations of the jet. Observing forbidden lines and assuming a projected plane of the sky velocity (150\kms) and length ($1\farcs25$), \cite{1995Hartigan} derived a jet mass-loss rate of $3.1\times10^{-8}$\msun\,yr$^{-1}$. We can update this value using the inclination of the outflow ($i=19\degr$), the line-of-sight blue-shifted jet velocity of 95\kms \citep[taking into account the cloud velocity and the observations of][]{1994Hirth}, the blue-shifted jet length from the HST observations ($\sim2\arcsec$) and the equation A10 from Appendix A.1 in \cite{1995Hartigan}. This way we obtain a jet mass-loss rate $\dot{M}_{jet}=4.3\times10^{-9}$\msun\,yr$^{-1}$. We can also make a rough estimate of the jet mass-loss rate assuming that it is a fraction of the accretion rate of the disk \citep[$\dot{M}_{jet}\simeq0.1\dot{M}_{acc}$, see e.g.,][]{2005Pudritz}. $\dot{M}_{acc}$ has been estimated as $1.4\times 10^{-7}$\msun\,yr$^{-1}$ and $3.0\times10^{-8}$\msun\,yr$^{-1}$ \citep{1998Gullbring,2017AlonsoMartinez}, yielding jet mass-loss rates ranging $3\times10^{-9}-10^{-8}$\msun\,yr$^{-1}$, in good agreement with our new estimate for the Hartigan et al. observations. Summarizing, the average mass load rate from the outflow shells is a factor $5-20$ larger than the jet mass-loss rate. In the meantime, its rate of injected momentum is comparable to that of the optical jet ($4\times10^{-7}$\msun\kms\,yr$^{-1}$). All this may suggest that the outflow shells are mainly comprised by swept-up gas from the original envelope/cloud, with only a small fraction of material incorporated from the wind directly launched from the disk. The momentum of this wind is apparently conserved when sweeping up the CO shells.

\begin{figure}
\includegraphics[angle=0, scale=0.48]{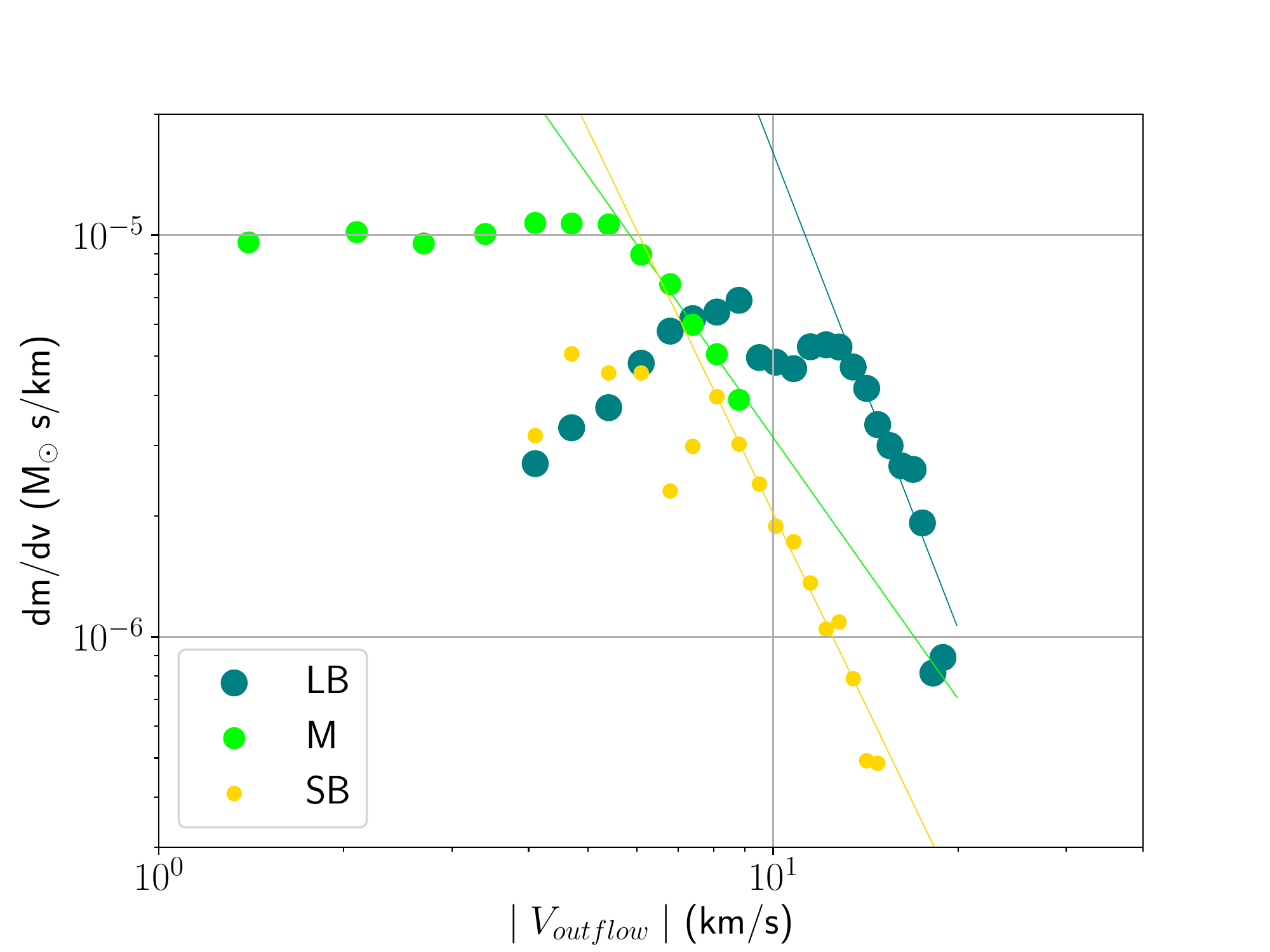}
\caption{\co shell masses per unit velocity interval as a function of the outflow velocity (corrected for inclination) for each of the three outflow shells. Power-law fits use only data corresponding to larger outflow velocities.  
} 
\label{fmasses} 
\end{figure}

Figure \ref{fmasses} shows the mass distribution $dm/dV$ within the shells as a function of velocity corrected by the outflow inclination. Symbols with three different colors designate the three outflow shells. An usual analysis of a mass spectrum diagram such as this, consists in perform power-law fits of the form $v_{outflow}\propto v^{-\gamma}$. We make individual fits to each of the three shells, using only masses corresponding to larger outflow velocities that show a clear power-law behavior. Table \ref{tglobal} collects the resulting slopes for each of the three fits. In the case of ambient shells driven by wide-angle winds the expected slope of the mass-velocity relationship is about 2 \citep[e.g.,][]{1992Masson,1996Li,1999Matzner}. Inclination effects \citep[e.g.,][]{2001Lee} and most importantly, deviations from the canonical angle dependence on the wind thrust and the ambient density distribution \citep{1999Matzner} can modify these values. In the case of DO\,Tauri's outflow, the slopes of the LB and SB shells are steeper than the slope of the M shell. The overall shape of the mass-velocity relations for the three shells matches quite well the expectations for the wide-angle wind scenario \citep[which are not very different from those of the jet-driven bowshock model,][]{1997Zhang,2003Downes}.

The resulting mass, momentum and energy of the blue-shifted lobe of the DO\,Tauri outflow are in general lower than those derived for some Class\,0 or Class\,I molecular outflows \citep[$\sim10^{-3}$\msun, $10^{-3}-0.1$\msun\kms, $10^{41}-10^{43}$\,erg][]{2003Arce,2004Arce,2014Lumbreras,2006Kwon,2014Zapata,2015Zapata,2016Zhang}. However they agree better with the characteristics of the outflow from the Class\,II HH\,30 \citep[$M=1.7\times10^{-5}$\msun, $\dot{M}=8.9\times10^{-8}$\msun\,yr$^{-1}$,][]{2018Louvet}. Thus, DO\,Tauri's outflow mass and energetics roughly lie within the range of other young stellar outflow values, supporting the outflow interpretation for the nested ringed structures in the surroundings of DO\,Tauri.

\begin{deluxetable}{lcccc} 
\tablewidth{0pt}
\tablecolumns{5}
\tabletypesize{\scriptsize}
\tablecaption{Physical properties of the three sets of rings}
\tablehead{
\colhead{} & \colhead{SB} & \colhead{M\tablenotemark{\dag}
} & \colhead{LB}  & \colhead{Total}
}
\startdata
Mean geometrical radius (au) & 195-240 & 280-350 & 640-850 & \nodata\\
Distance from DO\,Tauri (au)\tablenotemark{*} & 530-1025 & 190-910 & 1140-3680 & \nodata \\
T$_{dyn}$ (yr)  & 460$\pm70$ & 700$\pm$100 &  1090$\pm$90 & 1090$\pm$90 \\
M ($10^{-5}$\msun) & 2.6 & 4.0 & 5.9 & 12.5 \\
P ($10^{-4}$\msun\kms) & 2.1 & 2.5 & 6.5 & 11.1 \\
E (10$^{40}$ erg) & 1.8 & 1.6 & 7.6 & 10.9  \\
$\dot{M}$($10^{-8}$\msun\,yr$^{-1}$) & 5.6 & 5.7 & 5.4 & 11.4 \\
$\dot{P}$($10^{-7}$\msun\kms\,yr$^{-1}$) & 4.6 & 3.6 & 6.0 & 10.2 \\
$\gamma$ & 3.2$\pm$0.3 & 2.2$\pm$0.4 & 3.9$\pm$0.7 & \nodata\\
\enddata 
\tablenotetext{*}{Corrected by outflow inclination ($z=d_{observed}/\sin{(i)}$).}
\tablenotetext{\dag}{Magnitudes for the M shell are estimated taking into account only emission with $V_{outf}<-4$\kms}
\label{tglobal}
\end{deluxetable}

\subsection{Launching region}\label{s_launching}
The three CO shells of dragged gas have remarkably coherent and regular shapes several hundreds of au far away from the central source. In order to form structures with this degree of cohesion and shape, the launching zone of the outbursting wind or jet entraining these shells should be causally connected during the outburst period. Following \cite{2019Zhang} we constrain the dimensions of the launching area of the outbursts that entrained the outflow shells. We estimate the interval between the production of the SB and LB outflow shells in $\sim600$\,yr. Part of the M shell was intertwined between these two so we choose a shell interval time of 300\,yr. Assuming the ejection outburst lasts for just a fraction of the interval between the dynamical age of the shells (consider $t_{outburst}\simeq0.2-0.3\times\Delta t_{shell}$ as an upper limit to produce a coherent shell) we derive a length scale $c_s\cdot\Delta t_{outburst}$ using the disk sound speed $c_{s}=0.24\sqrt{T/10}$ expressed in \kms. Although this procedure is very approximate, by taking a temperature of 100\,K appropriate for the inner part of the disk, we obtain an upper limit for the size of the wind/jet outburst launching region of $<10-15$\,au.

\section{Conclusions} \label{sec:conclusions}
We presented new 1.3\,mm continuum and \co ALMA observations, and [SII] HST archival observations toward the DO\,Tauri young stellar object. The DO\,Tauri 45\,au disk is oriented north-south and is apparently close to face-on ($i=19\degr$). It shows a velocity gradient with this same orientation indicative of rotation. A rough estimate assuming a Keplerian rotation pattern gives a central mass of 1-2\msun.

The molecular environment of the region appears in the ALMA velocity cube as a series of nested ring structures, separated in radial velocity. The rings are spatially aligned with the location of DO\,Tauri's disk (lying along a line with PA$=253.6\degr$) and the optical jet. We identify the individual rings and fit them with ellipses deriving their size, orientation and inclination. We found that there are three sets of rings with separate sizes (SB  have radii $<260$\,au, LB  radii $>560$\,au and M have radii between these two). They show an increase of their radial velocity as a function of their distance to the star.

Based on the alignment of the rings with the jet, their kinematics and the high inclination of the young stellar disk, we interpret these ringed structures as part of a molecular outflow seen mostly pole-on (close to the line of sight), and therefore revealing an outflow from a different perspective. Instead of the typical parabolic biconical shape, DO\,Tauri's outflow is seen as a collection of elliptical rings (possibly the cavity walls) displayed at different radial velocities. After correcting for the line-of-sight inclination ($i=19\pm9\degr$) we derive the dimensions of the blue-shifted outflow (4070\,au$\times$1960\,au), its opening angle ($27\degr$) and its deprojected velocity (18\kms). From the observations we identified three outflow shells. Knowing the outflow inclination and adopting the wide-angle wind model of \cite{2000Lee}, we could derive the dynamical ages of the SB, M  and LB shells lie around 460\,yr, 700\,yr and 1090\,yr. These shells may correspond to entrainment by separate wind/jet outbursts. The total dynamical time of the outflow is about $1090\pm90$\,yr. We also estimated the mass ($1.3\times10^{-4}$\msun), the momentum ($1.1\times10^{-3}$\msun\kms) and the energy ($1.1\times10^{41}$\,erg) of the blue-lobe of the outflow and those of the individual shells (Table \ref{tglobal}).
The average mass load rate of the three outflow shells is about $5.5\times10^{-8}$\msun\,yr$^{-1}$ and that of the rate of injected momentum is $4.7\times10^{-7}$\msun\kms\,yr$^{-1}$. A comparison between these quantities and those of the optical jet indicates that only a fraction of the outflow material comes directly from the disk wind and the jet carry enough momentum flux to drive each of the shells.
All the measurements agree with the outflow hypothesis within the wide-angle wind model, but a jet-driven bowshock scenario can not be discarded, since the trend in velocity vs distance from source can be better explained by this model. Finally, we make a rough estimate of the dimensions of the wind/jet launching zone using the estimated elapsed time between the creation of each outflow shell. Our findings suggest a narrow ($<15$\,au) launching zone. 


New observations with deeper sensitivity will surely help to dig further into the complex molecular environment around DO\,Tauri offering a unique perspective of this phenomenon.        

\acknowledgments
We are very grateful to the anonymous referee for all the comments and suggestions that
helped a lot to improve the text and contents of this work. 

MFL has received funding from the European Union's Horizon 2020 Research and Innovation Programme under the Marie Sk\l{}odowska-Curie grant agreement No 734374

This paper makes use of the following ALMA data:
ADS/JAO.ALMA\#2016.1.01042.S. ALMA is a partnership of
ESO (representing its meMRr states), NSF (USA) and NINS (Japan), together
with NRC (Canada), NSC and ASIAA (Taiwan), and KASI (Republic of Korea), in
cooperation with the Republic of Chile.  The Joint ALMA Observatory is
operated by ESO, AUI/NRAO and NAOJ.  LAZ and LFR acknowledge financial
support from DGAPA, UNAM, and CONACyT, M\'exico. 

It is also partially based on observations made with the NASA/ESA Hubble Space Telescope, 
and obtained from the Hubble Legacy Archive, which is a collaboration between the Space 
Telescope Science Institute (STScI/NASA), the Space Telescope European Coordinating Facility 
(ST-ECF/ESA) and the Canadian Astronomy Data Centre (CADC/NRC/CSA).

\facility{ALMA, HST}
\software{DS9 \citep{2003Joye}, CASA \citep[v4.7.0 and v5.4.0][]{2007McMullin}, IRAF \citep{1986Tody}, Astropy \citep{2013Astropy}, Karma \citep{1995Gooch}, PDL \citep{1997Glazebrook}}

\newpage
\bibliography{biblio}

\appendix \label{sec:appendix}

\startlongtable 
\begin{deluxetable}{cccccccc} 
\tablewidth{0pt}
\tablecolumns{8}
\tabletypesize{\scriptsize}
\tablecaption{Results of ellipse fitting in the \co velocity cube}
\phs
\tablehead{
\colhead{Fit} & \colhead{Ellipse set} & \colhead{V$_{rad}$} & \colhead{RA offset} & \colhead{Dec offset} & \colhead{B$_{maj}$} & \colhead{B$_{min}$} & \colhead{PA} \\
\colhead{} & \colhead{} & \colhead{(\kms)}  & \colhead{($\arcsec$)} & \colhead{($\arcsec$)} & \colhead{($\arcsec$)}  & \colhead{($\arcsec$)} & \colhead{($\degr$)} 
}
\startdata
\cutinhead{Set of Large Blue-shifted Rings -- LB }
A & LB  & -12.1	& -9.4$\pm0.2$ & -2.3$\pm0.2$ & 6.6$\pm0.3$ & 5.7$\pm0.2$ & 113$\pm19$ \\
A & LB  & -11.5	& -9.3$\pm0.2$ & -2.7$\pm0.2$ & 7.0$\pm0.2$ & 5.6$\pm0.2$ & 79$\pm6$ \\
A & LB  & -10.8	& -9.4$\pm0.2$ & -2.5$\pm0.2$ & 7.1$\pm0.2$ & 5.8$\pm0.2$ & 88$\pm8$ \\
A & LB  & -10.2	& -9.2$\pm0.2$ & -2.9$\pm0.2$ & 6.9$\pm0.2$ & 6.0$\pm0.2$ & 77$\pm8$ \\
A & LB  & -9.6	& -8.7$\pm0.3$ & -2.8$\pm0.3$ & 7.0$\pm0.3$ & 5.5$\pm0.3$ & 62$\pm7$ \\
A & LB  & -8.9	& -8.8$\pm0.2$ & -2.6$\pm0.2$ & 7.2$\pm0.2$ & 6.0$\pm0.3$ & 71$\pm11$ \\
A & LB  & -8.3	& -8.1$\pm0.3$ & -2.5$\pm0.3$ & 6.6$\pm0.3$ & 5.7$\pm0.2$ & 70$\pm15$ \\
A & LB  & -7.7	& -7.6$\pm0.2$ & -2.3$\pm0.2$ & 6.2$\pm0.2$ & 5.7$\pm0.2$ & 63$\pm32$ \\
A & LB  & -7.0	& -7.1$\pm0.2$ & -1.8$\pm0.2$ & 5.9$\pm0.2$ & 5.6$\pm0.2$ & \nodata \\
A & LB  & -6.4	& -5.4$\pm0.2$ & -2.0$\pm0.3$ & 4.9$\pm0.2$ & 4.4$\pm0.3$ & \nodata \\
A & LB  & -5.8	& -5.7$\pm0.3$ & -1.8$\pm0.1$ & 5.0$\pm0.4$ & 4.9$\pm0.3$ & \nodata \\
A & LB  & -5.1	& -5.0$\pm0.2$ & -1.8$\pm0.1$ & 4.9$\pm0.2$ & 4.8$\pm0.2$ & \nodata \\
A & LB  & -4.5	& -4.7$\pm0.1$ & -1.8$\pm0.1$ & 5.0$\pm0.1$ & 4.9$\pm0.2$ & \nodata \\
A & LB  & -3.9	& -4.4$\pm0.1$ & -1.7$\pm0.1$ & 5.1$\pm0.1$ & 5.0$\pm0.1$ & \nodata \\
A & LB  & -3.2	& -4.3$\pm0.1$ & -1.6$\pm0.1$ & 5.1$\pm0.1$ & 5.1$\pm0.1$ & \nodata \\
A & LB  & -2.6	& -4.2$\pm0.2$ & -1.6$\pm0.1$ & 5.3$\pm0.2$ & 5.1$\pm0.2$ & \nodata \\
A & LB  & -1.9	& -3.8$\pm0.1$ & -1.5$\pm0.1$ & 5.4$\pm0.2$ & 5.2$\pm0.1$ & \nodata \\
A & LB  & -1.3	& -3.5$\pm0.2$ & -1.5$\pm0.1$ & 5.6$\pm0.2$ & 5.4$\pm0.1$ & \nodata \\
A & LB  & -0.7	& -3.2$\pm0.2$ & -1.3$\pm0.2$ & 5.8$\pm0.2$ & 5.6$\pm0.3$ & \nodata \\
A & LB  & 0.0	& -3.0$\pm0.2$ & -1.2$\pm0.2$ & 6.1$\pm0.2$ & 5.8$\pm0.2$ & \nodata \\
A & LB  & 0.6	& -3.1$\pm0.2$ & -1.1$\pm0.2$ & 6.3$\pm0.3$ & 5.9$\pm0.2$ & 153$\pm14$ \\
A & LB  & 1.2	& -3.6$\pm0.6$ & -0.7$\pm0.2$ & 6.3$\pm0.3$ & 6.0$\pm1.0$ & \nodata \\
A & LB  & 1.9	& -3.6$\pm0.8$ & -0.7$\pm0.2$ & 6.3$\pm0.3$ & 6.0$\pm0.4$ & \nodata \\
\cutinhead{Set of Medium Blue-shifted Ellipses -- MR} 
M & M & -2.6	& -2.3 & -0.6 & 2.2 & 1.8 & 135 \\
M & M & -1.9	& -2.2 & -0.7 & 2.3 & 1.8 & 135 \\
M & M & -1.3	& -2.0 & -0.6 & 2.3 & 1.9 & 135 \\
M & M & -0.7	& -1.7 & -0.5 & 2.3 & 1.9 & 138 \\
M & M & 0.0	& -1.8 & -0.5 & 2.4 & 2.2 & 125 \\
M & M & 0.6	& -1.2 & -0.6 & 2.2 & 2.0 & 132 \\
M & M & 1.2	& -1.0 & -0.4 & 2.4 & 1.9 & 152 \\
M & M & 1.9	& -1.0 & -0.5 & 2.5 & 2.0 & 152 \\
M & M & 2.5	& -1.1 & -0.6 & 2.6 & 2.2 & 156 \\
M & M & 3.1	& -0.9 & -0.8 & 3.0 & 2.1 & 168 \\
M & M & 3.8	& -1.0 & -0.6 & 2.8 & 2.1 & 179 \\
M & M & 4.4	& -0.7 & -0.4 & 2.9 & 1.9 & 163 \\
\cutinhead{Set of Medium Red-shifted Ellipses -- MR} 
M & M & 6.9	& -1.6 & -0.5 & 3.0 & 1.5 & 167 \\
M & M & 6.9	&  0.5 & -0.6 & 3.0 & 1.5 & 167 \\
M & M & 7.6	& -0.9 & -0.6 & 3.1 & 2.0 & 158 \\
M & M & 7.6	&  0.1 & -0.5 & 3.1 & 2.0 & 158 \\
M & M & 8.2	& -0.9 & -0.6 & 3.0 & 2.0 & 158 \\
\cutinhead{Set of Small Blue-shifted Ellipses -- SB } 
A & SB  & -8.3	& -2.42$\pm0.07$ & -0.75$\pm0.07$ & 1.65$\pm0.08$ & 1.46$\pm0.09$ & 153$\pm9$ \\
A & SB  & -7.7	& -2.50$\pm0.06$ & -0.85$\pm0.1$ & 1.6$\pm0.1$ & 1.6$\pm0.1$ & \nodata \\
A & SB  & -7.0	& -2.61$\pm0.06$ & -0.98$\pm0.09$ & 1.6$\pm0.1$ & 1.39$\pm0.09$ & 36$\pm12$ \\
A & SB  & -6.4	& -2.58$\pm0.07$ & -0.70$\pm0.08$ & 1.6$\pm0.1$ & 1.4$\pm0.1$ & 29$\pm23$ \\
A & SB  & -5.8	& -2.57$\pm0.08$ & -0.75$\pm0.07$ & 1.61$\pm0.08$ & 1.44$\pm0.07$ & 12$\pm33$ \\
A & SB  & -5.1	& -2.57$\pm0.06$ & -0.75$\pm0.07$ & 1.55$\pm0.09$ & 1.42$\pm0.08$ & 152$\pm24$ \\
A & SB  & -4.5	& -2.51$\pm0.05$ & -0.72$\pm0.06$ & 1.61$\pm0.09$ & 1.5$\pm0.1$ & 145$\pm16$ \\
A & SB  & -3.9	& -2.44$\pm0.06$ & -0.69$\pm0.06$ & 1.7$\pm0.1$ & 1.48$\pm0.09$ & 133$\pm10$ \\
A & SB  & -3.2	& -2.45$\pm0.07$ & -0.61$\pm0.05$ & 1.72$\pm0.07$ & 1.62$\pm0.05$ & \nodata \\
A\tablenotemark{(a)} & SB  & -2.6	& -2.4$\pm0.2$ & -0.7$\pm0.2$ & 1.6$\pm0.2$ & 1.5$\pm0.2$ & \nodata \\
M\tablenotemark{(a)} &	SB  &	-2.6 &	-2.2 &	-0.5 &	1.6 &	1.5 &	\nodata \\
M &	SB  &	-1.9 &	-2.2 &	-0.5 &	1.6 &	1.6 &	\nodata \\
M &	SB  &	-1.3 &	-2.1 &	-0.4 &	1.7 &	1.6 &	133 \\
M &	SB  &	-0.7 &	-2.1 &	-0.3 &	1.7 &	1.6 &	128 \\
M &	SB  &	0.0 &	-2.0 &	-0.5 &	1.8 &	1.6 &	79 \\
M &	SB  &	0.6 &	-1.6 &	-0.2 &	1.6 &	1.4 &	135 \\
M &	SB  &	1.2 &	-1.6 &	0.1 &	1.6 &	1.3 &	141 \\
M &	SB  &	1.9 &	-1.4 &	-0.3 &	1.6 &	1.2 &	159 \\
\enddata 
\tablecomments{Uncertainties for the manual fit are estimated in $\pm0\farcs3$ (3 pixels) for the RA and Dec coordinates, $\pm0\farcs3$ for the Bmaj and Bmin semi-axis of the ellipses, and $\pm15\degr$ for the PA of the ellipses. These uncertainties were estimated from pushing the fit on several ellipses up to the limit of reasonable ellipse boundaries and conservatively adding $0\farcs1$ to the errors in position and sizes and $5\degr$ to the PAs. We do not quote the PA of the ellipse when the eccentricity $\sqrt{1-(Bmin/Bmaj)^2}\leq$0.35 }
\tablenotetext{(a)}{This ring is fitted automatically and manually for comparison of the results.}
\label{trings}
\end{deluxetable}

\begin{figure*}
\centering
\includegraphics[angle=0, scale=0.65]{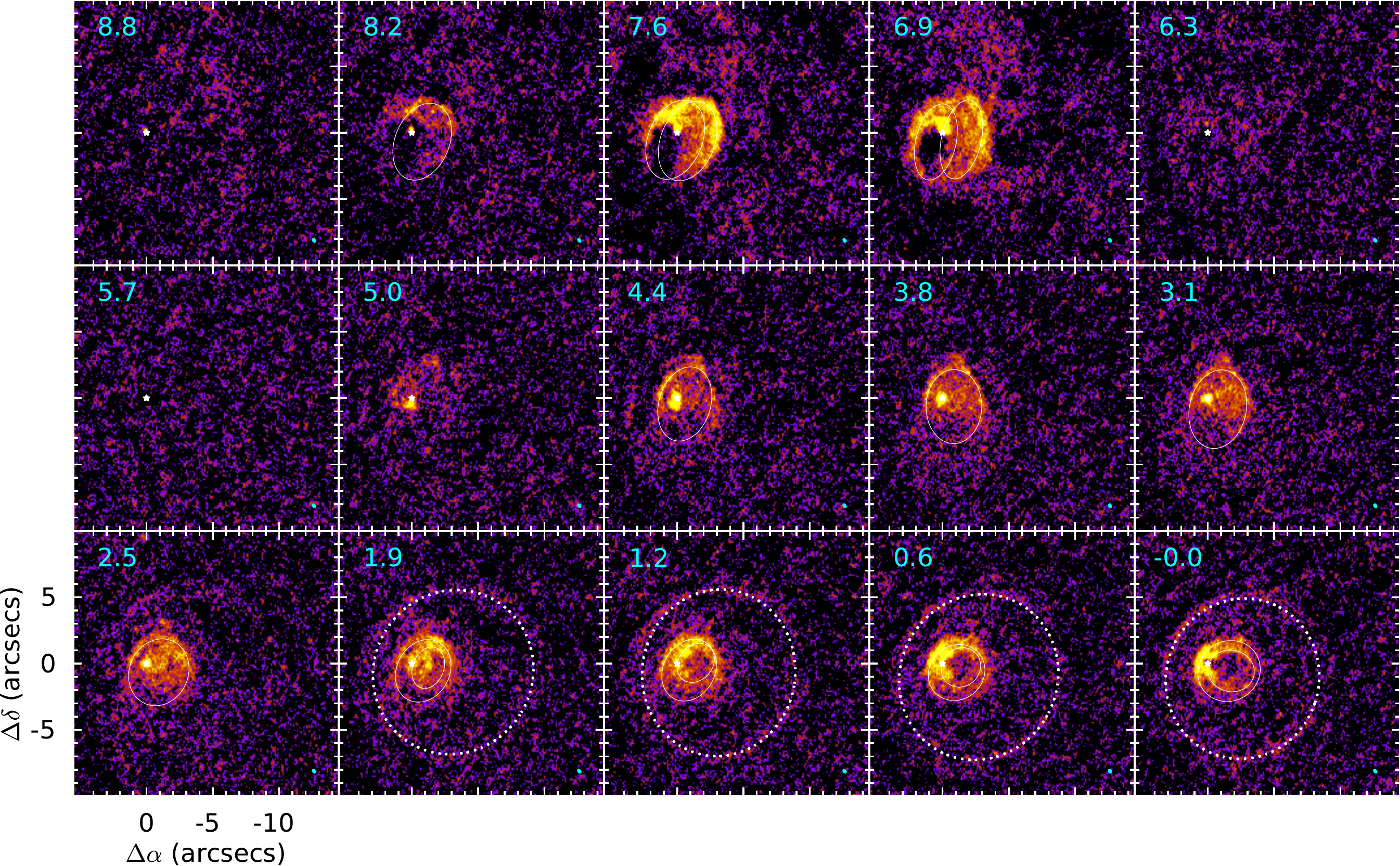}
\caption{Color scale image of the ALMA CO\,[J=2$\rightarrow$1] radial velocity channels from 8.8\kms to 0.0\kms. The position of the 1.3\,mm continuum peak is marked with a white star. The white dotted ellipses represent the fits carried out automatically, while the white solid ellipses show fits carried out manually (section \ref{sec:fits}). The velocity is indicated in the upper left corner and the synthesized beam is represented as a cyan ellipse in the bottom right corner of each channel respectively. The emission from the channels at the cloud velocity is filtered by the interferometer.
 } 
\label{large_red} 
\end{figure*}

\begin{figure*}
\centering
\includegraphics[angle=0, scale=0.65]{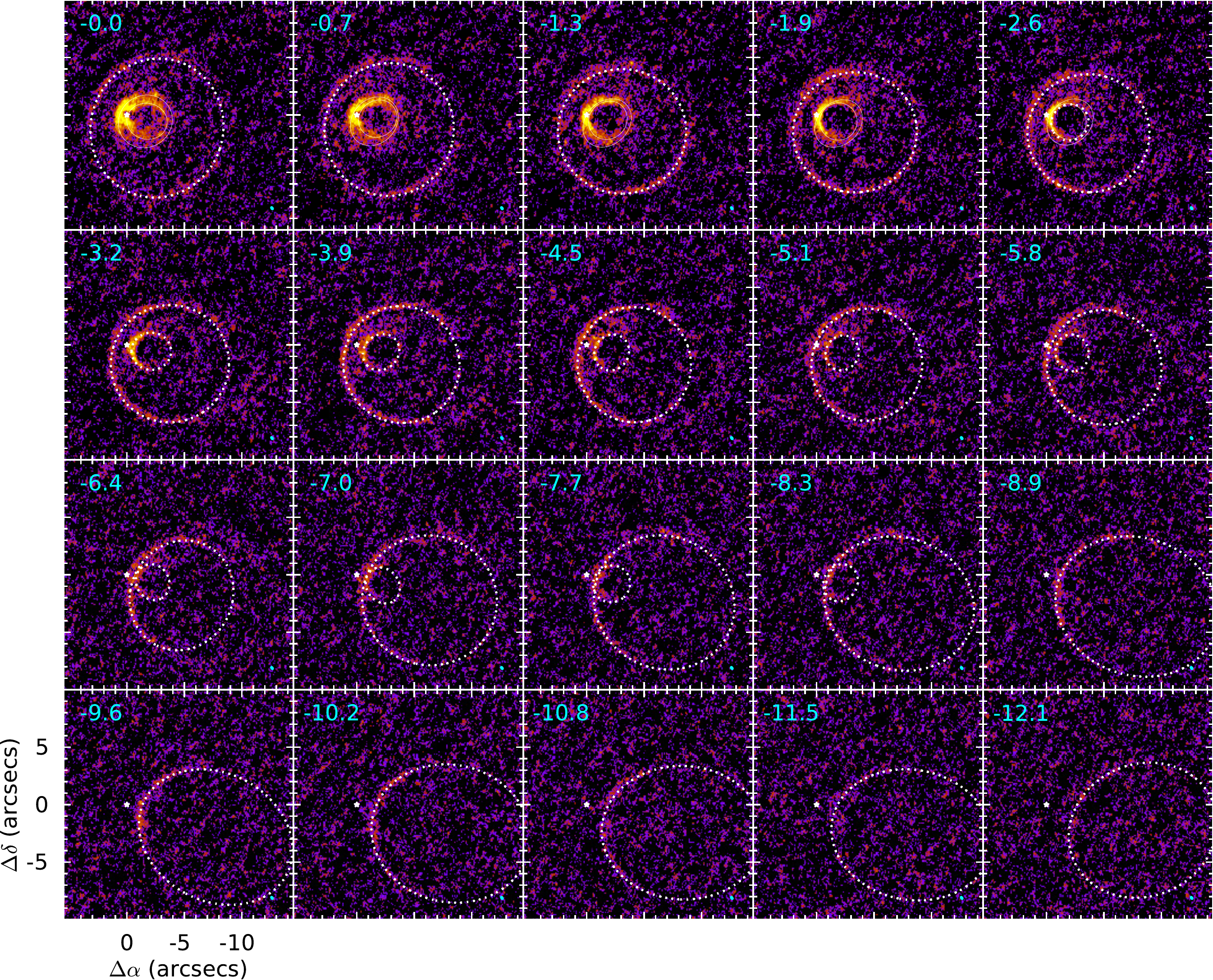}
\caption{Same as Figure \ref{large_red} but for the radial velocity interval between 0.0\kms and -12.1\kms.
 } 
\label{large_blue} 
\end{figure*}

\end{document}